\documentclass[a4paper,fleqn,usenatbib,useAMS]{mnras}

\usepackage{graphicx, natbib,epsfig,amsmath,amssymb, mathrsfs, txfonts}
\usepackage[latin9]{inputenc}
\usepackage{tikz}
\usetikzlibrary{shapes,arrows}

\usepackage[T1]{fontenc}
\usepackage{ae,aecompl}
\usepackage{array,multirow,makecell}

\usepackage{newtxtext,newtxmath}

\title{\textit{High-contrast at small separation: II. Impact on the dark hole of a realistic optical setup with two deformable mirrors.}}

\author[M. Beaulieu et al]{{M. Beaulieu$^{1}$\thanks{Contact e-mail: \href{mailto:mathilde.beaulieu@oca.eu}{mathilde.beaulieu@oca.eu}},  P. Martinez$^{1}$, L. Abe$^{1}$, C. Gouvret$^{1}$, P. Baudoz$^{2}$, R. Galicher$^{2}$}
\\
$^{1}$Universit\'e C\^{o}te d'Azur, Observatoire de la C\^{o}te d'Azur,, CNRS, Laboratoire Lagrange, France \\
$^{2}$LESIA, Observatoire de Paris, PSL Research University, CNRS, Sorbonne Universités, UPMC Univ. Paris 06, Univ. Paris Diderot, Sorbonne Paris Cité, France}

\date{Last updated 2015 May 22; in original form 2013 September 5}

\pubyear{2020}

\begin{document}
\label{firstpage}
\pagerange{\pageref{firstpage}--\pageref{lastpage}}
\maketitle

\begin{abstract}
Futurfdse large space or ground-based telescopes will offer resolution and sensitivity to probe the habitable zone of a large sample of nearby stars for exo-Earth imaging. To this aim, such facilities are expected to be equipped with a high-contrast instrument to efficiently suppress the light from an observed star to image these close-in companions. These observatories will include features such as segmented primary mirror, secondary mirror, and struts, leading to diffraction effects on the star image that will limit the instrument contrast. To overcome these constraints, a promising method consists in combining coronagraphy and wavefront shaping to reduce starlight at small separations and generate a dark region within the image to enhance exoplanet signal. We aim to study the limitations of this combination when observing short orbit planets. Our analysis is focused on SPEED, the Nice test-bed with coronagraphy, wavefront shaping with deformable mirrors (DM), and complex telescope aperture shape to determine the main realistic parameters that limit contrast at small separations. We develop an end-to-end simulator of this bench with Fresnel propagation effects to study the impact of large phase and amplitude errors from the test-bed optical components and defects from the wavefront shaping system on the final image contrast. We numerically show that the DM finite stroke and non-functional actuators, coronagraph manufacturing errors, and near focal-plane phase errors represent the major limitations for high-contrast imaging of exoplanets at small separations. We also show that a carefully-defined optical setup opens the path of high-contrast at small separation.

\end{abstract}
%



\begingroup
\let\clearpage\relax
\endgroup
\newpage

\section{Introduction}
As for today, more than \mbox{4 000} exoplanets have been discovered\footnote{from www.exoplanets.eu}, and only \mbox{$\thicksim$40} have been directly detected with masses greater than Jupiter mass, enlightening the difficulty to reveal Earth-like exoplanets in the habitable zone (defined as the liquid water zone). Direct detection enables detailed characterisation and in particular remote sensing of their atmospheres, opening the path of searching for habitability features.
Space-based telescopes, limited in size but free of atmospheric turbulence, should reach the contrast level required to image the reflected light of habitable planets around bright stars (A, F, or G types). Thermal emission of Earth-like planets should be directly imaged with the ground-based extremely large telescopes (ELTs, large primary mirror but affected by the Earth's atmosphere) around M-stars (fainter stars with the closer habitable zone). 
Despite atmospherics considerations, those two complementary methods share the common challenge of developing instruments providing high contrast at small angular separations. The large UV-optical-infraRed (LUVOIR, \citealt{BolcarCrookeHylanEtAl2018,PueyoStarkJuanolaaEtAl2019,JuanolaZimmermanPueyoEtAl2019}) and the habitable exoplanet observatory (HABeX, \citealt{KristMartinKuanEtAl2019, GaudiSeagerMennessonEtAl2018}) spatial mission concepts, for instance, aim to achieve a contrast ratio of $10^{-10}$ at \mbox{2 $\lambda$/D} (with $\lambda$ defining the wavelength and $D$ the telescope aperture diameter) to detect Earth-like planets around bright stars. Their ground-based counterparts are expected to detect Earth-like planets around M-stars if instruments achieve contrast ratios of $10^{-8}$ at \mbox{1 $\lambda$/D} (\citealt{GuyonMartinacheCadyEtAl2012}).

High-contrast imaging instrumentation and observation techniques have to face various challenges: (1) unfriendly pupil shape (primary mirror central obscuration and segmentation, etc.) that degrades the contrast ratio, (2) small-angular-separation regime where a large amount of the on-axis point-spread function (PSF) is concentrated, (3) dynamic, static, and quasi-static aberrations due to environment instabilities (change of temperature, pressure, and gravity induce deformations in the structure and mirrors, vibrations, etc.) that significantly degrade the performance.  High-contrast imaging requires multiple-step corrections such as extreme-adaptive optics (ExAO, for ground-based observatories), non-common path aberration control (i.e., differential aberrations between sensing and science paths), diffraction suppression or coronagraphy, active optics (cophasing for segmented telescope and regular active optics for all telescopes, e.g., telescope adaptive mirrors) and science image post-processing.

Pillars of high-contrast imaging at small angular separations include sophisticated techniques, amongst others:
\begin{enumerate}
\item coronagraphic devices adapted to obstructed/segmented pupils and to small separation regime (e.g., vortex coronagraph -- \citealt{MawetRiaudAbsilEtAl2005, FooPalaciosSwartzlander2005}, phase-induced amplitude apodization -- \citealt{GuyonMartinacheBelikovEtAl2010}) but at the cost of high sensitivity to aberrations;
\item deformable mirrors (DM) technology for wavefront shaping. Rather than the wavefront control which flattens the wavefront errors from imperfect optics, the wavefront shaping is the process of creating a dark zone (the so-called dark hole) in the PSF by locally minimising the light in the focal plane. One limitation of wavefront shaping is the Fresnel propagation of phase aberrations (also called the "Talbot effect"): at the DM plane, out-of-pupil optics create a mix of amplitude and phase errors that a single DM can only correct at the expense of losing at least half of the field in the science image (e.g. \citealt{BordeTraub2006, GiveonKernShaklanEtAl2007}). One way to deal with this effect is the use of at least two DMs to correct for both phase and amplitude over the full field;
\item quasi-static speckle calibration via post-processing, e.g., locally optimised combination of images (LOCI, \citealt{LafreniereMaroisDoyonEtAl2007}), observational strategies that take benefit from different behaviour of the planet and the star speckles, i.e., rotation angle (angular differential imaging, ADI, \citealt{MaroisDoyonRacineEtAl2005}), azimutal spectral dispersion (spectral differential imaging, SDI, \citealt{MaroisLafreniereDoyonEtAl2006}), spectrum (spectral deconvolution, SD, \citealt{SparksFord2002}) or polarisation (polarisation differential imaging, PDI, \citealt{KuhnPotterParise2001}), and coherent differential imaging (CDI) that uses modulation to determine the coherent and incoherent part of the field (i.e., \citealt{BordeTraub2006, GiveonKernShaklanEtAl2007, BaudozBoccalettiBaudrandEtAl2006})
\end{enumerate}

Current observing sequences with, for instance, GPI (\citealt{MacintoshGrahamPalmerEtAl2007}), SPHERE (\citealt{BeuzitFeldtDohlenEtAl2008}), or SCExAO (\citealt{GuyonMartinacheGarrelEtAl2010}) widely use reference PSF subtraction methods to improve high-contrast performance. However, those solutions are mostly inadequate at small angular separations where most of the starlight is concentrated, because of insufficient chromatic speckle elongation or field of rotation, etc. Improving the contrast performance at short separation imposes reducing the pupil plane wavefront errors with low spatial frequency to control the star energy distribution close to the optical axis in the science image. Speckle intensity in a coronagraphic image at close angular separation from the star is a highly non-linear function of wavefront errors which makes the situation very complex to tackle.

To date, the most advanced ground-based instruments (SPHERE, GPI, etc.) have yielded to the discovery of less than five new exoplanets. Improving the level of detection of these instruments and/or anticipating the performance of new instruments in the ELT area can be achieved by (i) improving the correction of non-common path aberrations, (ii) improving the attenuation at a smaller inner working angle (IWA), (iii) improving the ExAO correction temporal frequency and sensitivity, and finally, (iv) implementing sophisticated multi-DM wavefront shaping system dedicated to dark-hole generation.
Wavefront shaping systems have been tested in the laboratory (e.g \citealt{LawsonBelikovCashEtAl2013,MazoyerBaudozBelikovEtAl2019}) and on-sky (e.g \citealt{SavranskyMacintoshThomasEtAl2012,MartinacheGuyonJovanovicEtAl2014,BottomFemeniaHubyEtAl2016}) but are not yet routinely implemented on on-sky instruments. In this perspective, the understanding of limiting parameters for high-contrast imaging at small separations is crucial and timely and has led to the development of various laboratory test-beds and end-to-end simulations worldwide. 

While laboratory setups have been developed to evaluate high-contrast imaging instrument technologies and concepts (e.g., SPEED - the segmented pupil experiment for exoplanet detection -  \citealt{MartinezPreisGouvretEtAl2014}, HiCAT - the high-contrast imager for complex aperture telescopes - \citealt{NDiayeChoquetPueyoEtAl2013}, HCST - the high-contrast spectroscopy test-bed for segmented telescopes -  \citealt{MawetRuaneXuanEtAl2017}, DST - the decadal survey testbed - \citealt{GarrettCrillEtAl2019}, THD2 - \textit{très haute dynamique} - \citealt{BaudozGalicherPotierEtAl2018} and VODCA - the vortex optical demonstrator for coronagraphic applications - \citealt{JolivetPironHubyEtAl2014}), end-to-end numerical simulations are used to understand and predict high-contrast performance and limitations. The latter have been developed for instance for the WFIRST spatial mission (\citealt{KristNematiMennesson2016}), for LUVOIR (\citealt{JuanolaParramonZimmermanGroffEtAl2019}), or for generic high-contrast instruments (\citealt{BeaulieuAbeMartinezEtAl2017}). End-to-end simulators algorithms frequently include Fresnel propagator (\citealt{Krist2007a}) and/or energy minimisation algorithm for dark-hole generation (\citealt{GiveonKernShaklanEtAl2007, PueyoKayKasdinEtAl2009,RiggsRuaneSidickEtAl2018}). 

In particular, the end-to-end simulations developed in \citealt{BeaulieuAbeMartinezEtAl2017} aimed to determine the optimum wavefront control architecture for high-contrast imaging at small separations (around \mbox{1 $\lambda$/D}) using the combination of coronagraphy and wavefront shaping. The study assumed a generic high-contrast architecture with a perfect coronagraph, a monolithic circular aperture without any central obscuration nor spiders, etc., to assess the impact of the location of the two DMs on wavefront shaping when assuming the Fresnel propagation of standard aberrated optics. The objective of the study was to restrict the analysis to the intrinsic properties of the optics setup including polishing frequency distribution, relative beam size, the distance between optics, DMs optical location (in a collimated beam - out-of pupil plane or in a pupil plane - vs. converging beam) and separation, DMs properties (actuator number, etc.). The analysis has shown that high-contrast imaging at small separations with multi-DMs architecture requires large DMs separations. In particular, a significant performance dependence on the DM location, on the aberrations amount and power spectral density (PSD) power law and dark-hole size, have been demonstrated. 

The goal of the present paper is to provide a more realistic instrument setup design to go further in the analysis by assessing the relative impact of setup parameters (non-uniform source, residual pupil phasing aberrations, highly-aberrated optics, realistic deformable mirrors, and coronagraph, etc.) with a segmented and obstructed pupil. 

Ideally, such analysis is instrument-dependent and would require a case to case study. Nonetheless, for the sake of generality and to derive guidelines for high-contrast imaging development, we use the SPEED test-bed as a typical instrument for our study. The SPEED laboratory setup is specifically intended for high-contrast imaging at very small angular separations with an ELT-like telescope simulator. SPEED includes a segmented and centrally obstructed primary mirror, cophasing optics, a coronagraph designed for small-IWA with ELTs (a PIAACMC, phase-induced amplitude apodization complex mask coronagraph, \citealt{GuyonHinzCadyEtAl2014}), and a dual-DM wavefront shaping system. SPEED can be considered as representative of most of the current laboratory benches mentioned previously by sharing to some extent common optical system architectures, hardware, and/or observing strategies, though some objectives and parameters are different. The complexity of its optical design along with its numerous optics (25) and optics quality are typical to what current high-contrast instruments would incorporate apart from (i) the ExAO system, and (ii) a few sub-systems required to interface with the telescope that are not relevant in a laboratory environment (derotator, atmospheric dispersion compensator, etc.) are not taken into account. We use the sample of the SPEED set up, but we are interested in identifying fundamental limits for the entire field of exoplanet imaging (ground and space) thus our simulations are "turbulence free" and go to contrasts congruent with a possible future space telescope. The current paper focuses on the challenging small separation regime (less than \mbox{2 $\lambda$/D}), in order to assess practical limitations to high-contrast imaging.

The general assumptions used for the analysis (speckle pattern modelling, dark hole algorithm and numerical implementation) are described in section \ref{sec:assumptions}. Section \ref{sec:real} describes the SPEED test-bed and the realistic assumptions used for the analysis whereas section \ref{sec:res} shows the simulated effect on the contrast ratio performance. Finally, we provide a conclusion.

\section{General assumptions}
\label{sec:assumptions}
In this section, we describe the general assumptions on optics and we explain our numerical modelling methodology and dark-hole algorithm following the same formalism as for \citealt{BeaulieuAbeMartinezEtAl2017}.

\subsection{Speckle distribution modelling}
The diffraction pattern in the science image results from the pupil shape, the aberration all along the test-bed and the detector noises. For this analysis we have defined: (i) a segmented and obstructed pupil, which corresponds to the SPEED pupil and mimics the ELT features (see section \ref{sec:speed});
(ii) an optical setup with static aberrations. We do not treat here quasi-static aberration as we assume that the correction timescale is shorter than structural or thermal changes. The static aberration is simulated as followed: each optic is computed with random static aberrations defined by their total amount of aberration (in nm rms over the optic physical size) and their frequency distribution (power law of the power spectral density, PSD). We define each paraxial lens with standard optic qualities, i.e., with 5 nm rms aberration and a power law of the PSD in $f^{-3}$ (typical to current manufacturing errors). For statistical analysis, \mbox{128} phase realisations are defined per optic and the performance is computed for each of the 128 cases. The statistical validity of using 128 realisations has been verified (\citealt{BeaulieuAbeMartinezEtAl2017}).
We do not treat in this paper, detector noises (readout or current noises) or wavefront sensing errors.
\subsection{Dark hole algorithm}
\label{sec:dhalg}
The analytical approach, based on the energy minimisation, for generating a dark-hole in the science image is described in \citet{GiveonKernShaklanEtAl2007}, \citet{PueyoKayKasdinEtAl2009}, \citet{Groff2012} and \citet{Beaulieu2017}, and is defined by computing the total energy at the image plane when assuming a first DM at the pupil plane and a second DM at an out-of-pupil plane. 
We resume here the formalism for clarity.
We define: 
\begin{enumerate}
\item \textit{$E_0$} as the initial aberrated field with its amplitude \textit{A} and its phase \textit{$\varphi$}, 
\item \textit{C$_1$} as the linear operator from the pupil plane (where the first DM is located) to the focal plane, 
\item \textit{C$_2$} as the linear operator from the second DM (out-of-pupil plane) to the image plane,
\item \textit{C$_{12}$} as the linear operator from the first to the second DM plane,
\item \textit{a} as the DMs phases coefficients,
\item \textit{g} and \textit{h} as the influence functions of respectively the first and second DM.
\end{enumerate}
We assume that all the phases are small enough to approximate  \textit{$e^{i\varphi}$} by  \textit{$1+i\varphi$}, and that  \textit{$C_{12}[E_0.e^{i\phi_1}]$} can be written in the form of  \textit{$Ae^{i\phi}$} (and thus can be approximated by $A(1+i\phi)$).
The intensity inside the dark hole can be written as 
\begin{equation}
 \mathrm{I_{DH}= {^t}a \ M_0 \ a + 2 \ {^t}a  \ \Im(b_0) + d_0}, 
 \label{eq:idh}
 \end{equation}
\begin{align*}
 \mathrm{where} \quad & \mathrm{M_0= G^*G,} \\ 
 &  \mathrm{G=[ G_1, G_2],} \\
 & G_1= \left[ \begin{array}{ccc}
  & & \\
  & [C_1\{Ag_j\}]_i & \\
  & & 
  \end{array}
  \right], \\
  & G_2= \left[ \begin{array}{ccc}
  & & \\
  & [C_2\{Ah_j\}]_i & \\
  & & 
  \end{array}
  \right], \\
& b_0= \left[ \begin{array}{c}
  G_1^*\ C_1\{E_0\} \\
  G_2^*\ C_1\{E_0\} \\
    \end{array}
  \right], \\
& d_0=\langle C_{1}\{E_0\},C_{1}\{E_0\} \rangle.
\end{align*}
$M_0$ represents the system response to each DM poke, $b_0$ represents the interaction between the DM and the aberration, and $d_0$ is the initial intensity with aberrations and flat DMs ($a=0$).
The solution 
\begin{equation}
\mathrm{
\mbox{$a=-M_{0}^{-1}\Im(b_0)$},}
\label{eq:coeff}
\end{equation}
that represents the DM coefficients, minimises the energy inside the dark hole. Other algorithms such as electric field conjugation (EFC, \citealt{GiveonKernShaklanEtAl2007}) and the stroke minimisation method \citep{PueyoKayKasdinEtAl2009} optimise the contrast ratio and limit large stroke deviation. Because our model uses monochromatic light and assumes a perfect wavefront sensor, in our analysis, we do not handle large stroke deviation. 
We thus apply equation \ref{eq:coeff} without any stroke limitation.

\subsection{Dark hole numerical implementation}
The dark hole algorithm is implemented following section \ref{sec:dhalg} and is described in details in \citealt{BeaulieuAbeMartinezEtAl2017}. The interaction matrix $M_0$ is computed by first poking each DM actuator, then Fresnel-propagating the wavefront from the DM to the focal plane and finally recording the complex amplitude (we assume a perfect wavefront sensor). Because the DM surface solution is in phase (real DM stroke), we invert the real part of the matrix $M_0$. 

As illustrated in figure 9 of \citealt{BeaulieuAbeMartinezEtAl2017}, two iterative processes are necessary to optimise the DMs coefficients. The first one concerns the inversion of the matrix $M_0$ that can lead to diverging solution because of the presence of very low singular values. These are sorted out from the highest to the lowest values to exclude some of them. The algorithm starts with the \textit{n} first singular values from the singular value decomposition (and zeroing the remaining values) to compute the DMs coefficients. Iteratively, the algorithm includes the next $n$ singular values (previously zeroed), until the best contrast ratio is achieved. We empirically set the singular value threshold \textit{n} to 5 because it represents a fair compromise between the performance and the computational time. In theory, applying the DMs coefficients \mbox{$a_{th}=-M_{0}^{-1}\Im(b_0)$} leads to the nominal theoretical intensity \mbox{$I_{th}={^t}a_{th} \ M_0 \ a_{th} + 2 \ {^t}a_{th}  \ \Im(b_0) + d_0$}. However, small non-linearity of the optical operators $C_1$, $C_2$ and $C_{12}$ leads to a not optimal solution ($I_{DH} \neq I_{th}$) and needs a second iterative process to optimise the contrast until the theoretical intensity $I_{th}$ is reached.

The analysis presented in \citealt{BeaulieuAbeMartinezEtAl2017} has determined that high contrast at small separations requires large DMs distances from the pupil but also small dark hole size; we have thus defined the dark hole from 0.8 to \mbox{4 $\lambda/D$} to emphasise performance at very small separations. 
The code we use for the Fresnel propagation between each optical element is \textsc{proper} \citep{Krist2007a}. \textsc{proper} and the dark hole algorithm were written in \textsc{idl} but translated in \mbox{\textsc{c++}}, such that the computation of the 128 configurations are performed simultaneously in a data centre available at Observatoire de la C\^{o}te d'Azur to speed up the computational time (from one day to several hours for all the configurations).

\section{Realistic optical setup definition}
\label{sec:real}

This section introduces the optical setup architecture and the working hypothesis that will offer an adequate playground for our study. 

\subsection{Generic high-contrast imaging model}
Many worldwide high-contrast imaging test-beds uses several DMs for wavefront shaping (e.g., DST, HCST, HiCAT, SPEED, THD2). In particular, HiCAT, DST, HCST, and SPEED, incorporate a segmented and centrally-obstructed pupil in their telescope simulator and thus implement closed-loop cophasing optics. Despite different objectives and parameters (different field-of-view (FoV), coronagraphs, wavefront sensing methods, etc.), they share a common optical architecture consisting in \mbox{$\thicksim$ 25} optics, including the source module, a tip/tilt correction, a coronagraph, two (or more) DMs and off-axis parabolas (OAPs) to ensure the transition between pupil and focal planes.
In this context, the SPEED test-bed aims to achieve high contrast at very small separations in H-band with an ELT-like pupil,
perfectly adapted to the problem of determining realistic limitations of high-contrast imaging at very small separations. This work focuses on the SPEED test-bed (coronagraph, segmented/obstructed pupil, optical design, etc.) and the results can be applied to other high-contrast test-beds because they share most of the setup parameters.

\subsection{The SPEED test-bed}
\label{sec:speed}
\begin{figure*}
\centering
\includegraphics[width=1.7\columnwidth]{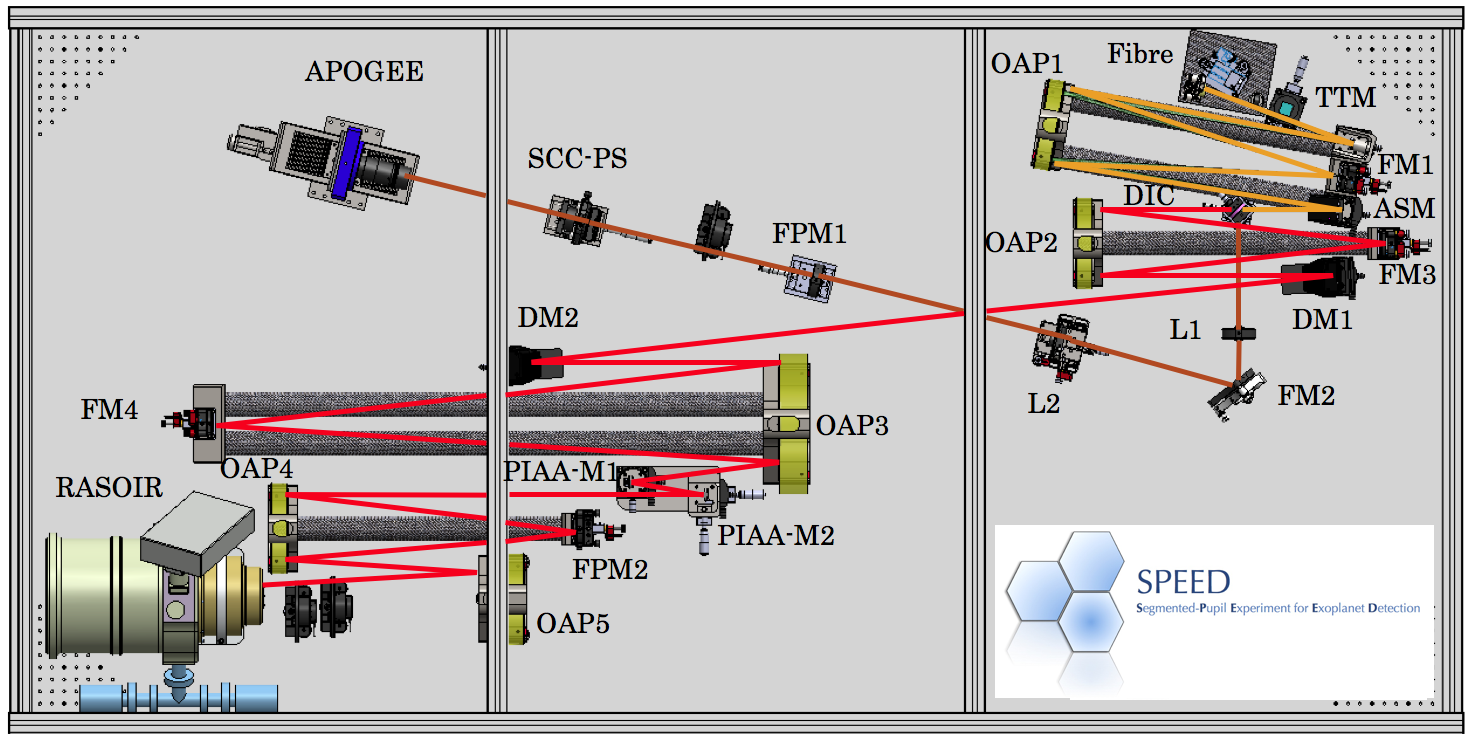}
\caption{SPEED test-bed. The common path (before the dichro\"{i}c) is shown in orange, the visible path is represented in brown and the NIR path is in red. The acronyms on the figure are : L for lens, OAP for Off-Axis Parabola, ASM for Active Segmented Mirror, DM for Deformable Mirror, FM for Flat Mirror, DIC for dichro\"{i}c, SCC-PS for Self-Coherent Camera-Phasing Sensor, FPM for Focal Plane Mask, PIAA-M1 and PIAA-M2 for the two PIAA mirrors for apodization, LS for Lyot Stop, APOGEE for the visible camera and RASOIR for the NIR camera. }
\label{fig:setup}
\end{figure*}

\begin{figure}
\centering
\includegraphics[width=.9\columnwidth]{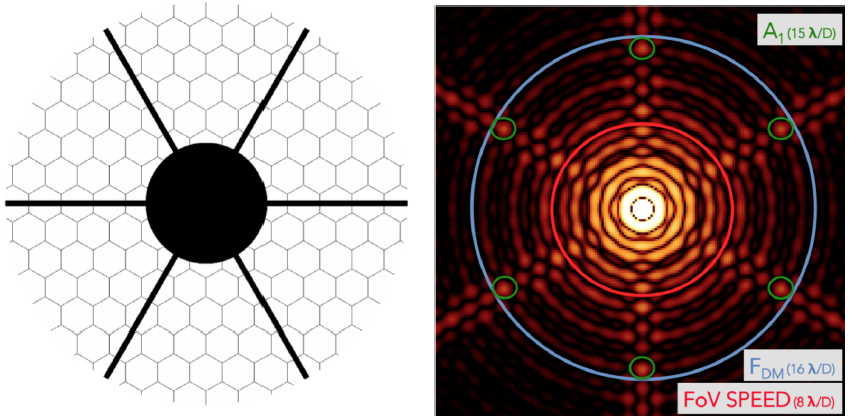}
\caption{SPEED pupil (left) and corresponding FoV (right)}
\label{fig:pup}
\end{figure}

The SPEED test-bed combines a source module and a telescope simulator (orange line), a dichroïc reflecting the visible light to cophasing optics (brown) and transmitting the near-infrared light toward wavefront shaping, coronagraphy, and science camera (red). The bench layout and hardware are illustrated in figure \ref{fig:setup}. The test-bed includes the following elements: 
\begin{enumerate}
\item \textit{a source module} made of a super-continuum light source feeding an optical fibre combined with a spherical mirror collimating the beam onto the tip-tilt mirror;
\item \textit{a tip-tilt mirror} to guarantee the stabilisation of the PSF on the coronagraph;
\item \textit{a telescope simulator} combining an active segmented mirror (ASM) comprising 163 segments, controlled in piston and tip/tilt, and an optical mask inserted into the beam on the tip-tilt mirror to simulate a large central obscuration (30\%) and 6 spiders separated by 60 degrees. The pupil, shown on the left of figure \ref{fig:pup}, is 7.7 mm in diameter.
The corresponding PSF is shown on the right of figure \ref{fig:pup}, illustrating the diffraction effects due to segmentation (green circles), the DMs cut-off frequency (blue circle), and the defined SPEED FoV (\mbox{8 $\lambda$/D}, red circle);
\item \textit{cophasing sensors} including the self-coherent camera phasing sensor (SCC-PS, \citealt{Janin-PotironMartinezBaudozEtAl2016}) and alternatively a Zernike-based phasing sensor (\citealt{Janin-PotironNDiayeMartinezEtAl2017});
\item \textit{a wavefront shaping system} combining two continuous facesheet DMs from Boston Micromachine\footnote{http://www.bostonmicromachines.com}. The DMs are made of \mbox{34 $\times$ 34} actuators with an inter-actuator pitch of \mbox{300 $\mu$m}. The two DMs are located at 1.5 and \mbox{0.2 m} on both sides of the pupil plane, to maximise the performance between 0.8 and \mbox{4 $\lambda$/D} (\citealt{BeaulieuAbeMartinezEtAl2017});
\item \textit{a PIAACMC} offering high-throughput and a small IWA of \mbox{1 $\lambda$/D}. The PIAACMC combines three elements: a lossless apodization with aspheric mirrors to weaken the Airy rings, a phase-shifting focal plane mask, and a Lyot stop that blocks the diffracted light. The design, specifications and manufacturing of the coronagraphic prototype is detailed in \citealt{MartinezPreisGouvretEtAl2014};
\item \textit{an infrared camera} operating at the wavelength of \mbox{1.65 $\mu$m} with a 1k by 1k Hawaii detector array.
\end{enumerate}

\subsection{Realistic simulations}
\label{sec:reals}
End-to-end simulators have been developed with various objectives and complementary analysis; we here focus on few of them. 
In \citealt{KristBelikovPueyoEtAl2011} (developed as part of the technology demonstration for exoplanet mission and applied to the WFIRST mission) the authors compare the performance of different coronagraphs in polychromatic light when taking into account the propagation of both phase and amplitude errors on each optic. On the other hand, in \citealt{BeaulieuAbeMartinezEtAl2017}, we statistically analyse the impact of phase errors on the performance in monochromatic light and at very small separations assuming a perfect coronagraph. Finally, in \citealt{JuanolaParramonZimmermanGroffEtAl2019} the authors determine the impact of the telescope aberration on the performance in polychromatic light for the LUVOIR mission assuming perfect instrument optics and an APLC coronagraph. 

In the current paper, we present the relative impact of setup parameters on the ability to efficiently control phase and amplitude to create dark holes at very small separations (around \mbox{1 $\lambda$/D}) when taking into account Fresnel propagation of standard optics errors but also specific and realistic optical setup parameters. We consider in our analysis three main categories of errors: (i) phase errors that are not taken into account in the nominal case: highly-aberrated optics (e.g., deformable mirrors windows), residual phase aberrations on the segmented pupil, or coronagraph realistic manufacturing errors; (ii) amplitude errors due to the source module such as segment reflectivity variation or missing segment in the pupil;
(iii) errors from the active correction system itself such as stroke limitation or non-functional actuator. 

\section{Numerical results}
\label{sec:res}

This section describes the impact of realistic parameters (previously defined in section \ref{sec:reals}) on the performance. The performance criteria is 
defined as the $5\sigma$ contrast ratio histogram computed for each of the 128 random realisations. It corresponds to the number of random realisations that achieves a given contrast ratio inside the defined dark hole (see section \ref{sec:nom} for illustration purpose). The numerical pupil diameter size is \mbox{225} pixels corresponding to the \mbox{ 7.7 mm} pupil diameter for a grid size of \mbox{1024} pixels. The simulation is monochromatic at the wavelength of \mbox{1.65 $\mu$m}. 

\subsection{Nominal case}
\label{sec:nom}
We define a nominal case as a comparison basis to assess the relative impact of each parameter. It corresponds to a 25 optics setup containing: (i) an obscured mask with spiders located onto a tip-tilt mirror, (ii) a perfectly co-phased segmented mirror (with 163 segments), (iii) a theoretical PIAACMC coronagraph, and (iv) two DMs (34x34 actuators) with infinite stroke and located at 1.5 and 0.2 m from the pupil plane. 

\begin{figure}
\begin{center}
\includegraphics[width=.4\columnwidth]{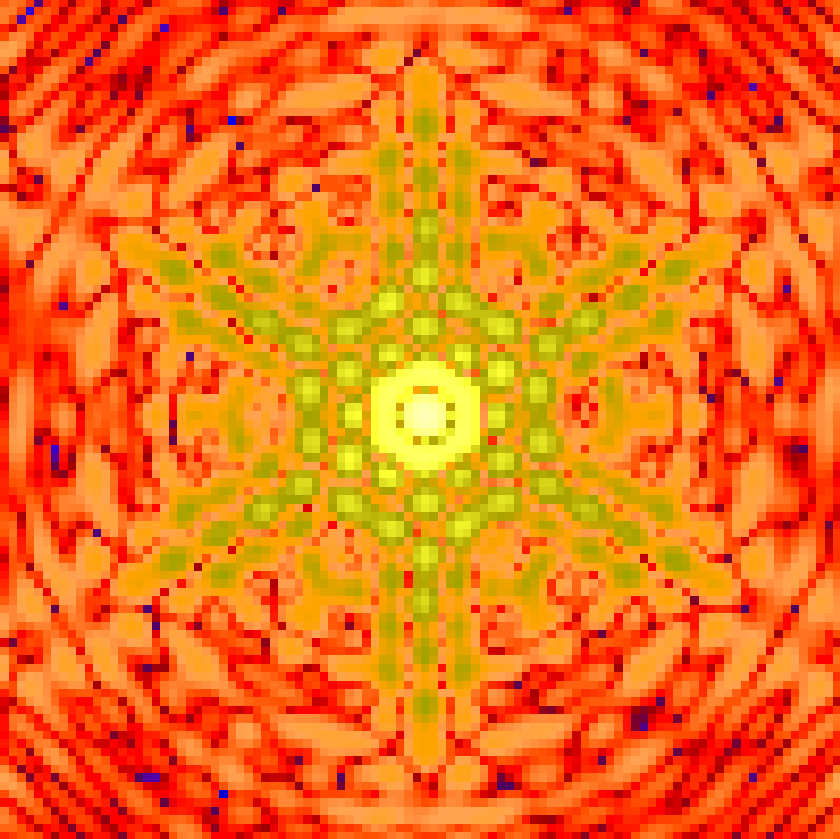}
\includegraphics[width=.4\columnwidth]{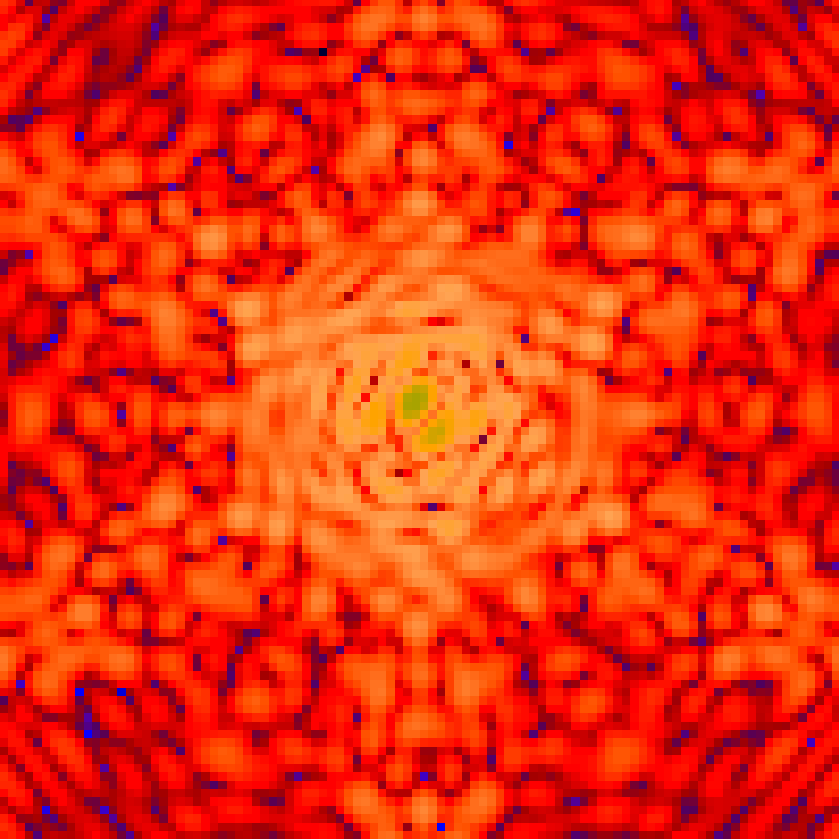}
\includegraphics[height=.4\columnwidth]{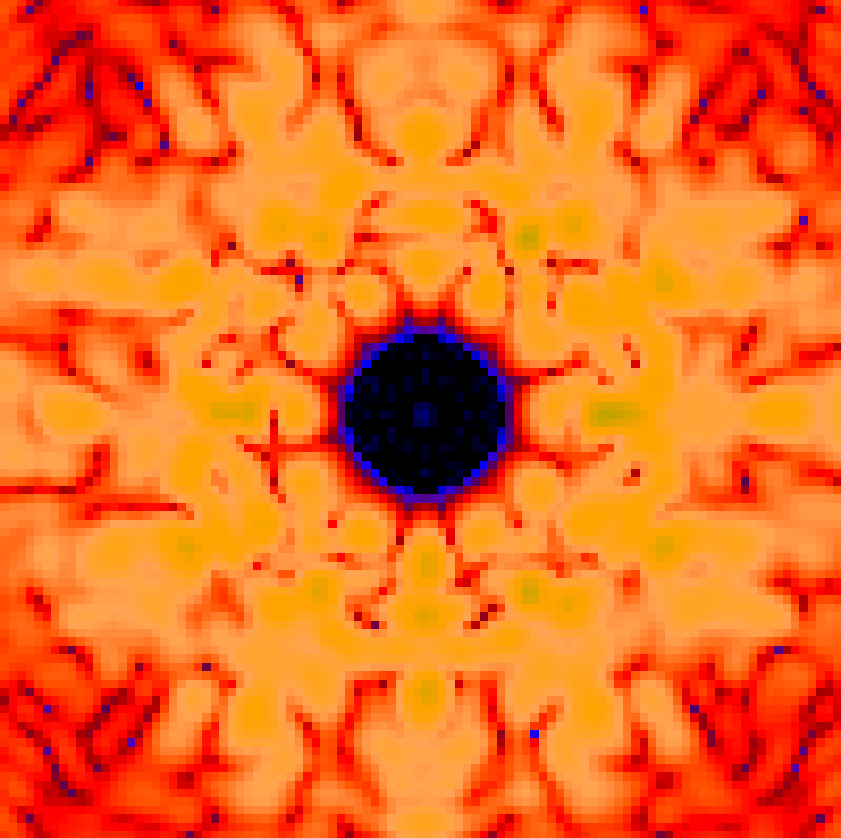}
\includegraphics[height=.4\columnwidth]{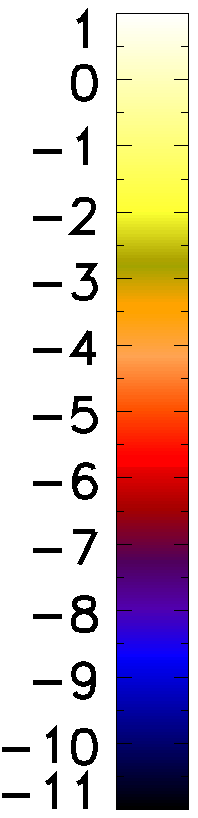}
\end{center}
\caption{Contrast ratio images in logarithmic scale before the coronagraph (top left) and after the coronagraph and before (top right) and after (bottom) the dark hole algorithm. The overall figure represents 44$\lambda$/D in size.}
\label{fig:res_ill}
\end{figure}

\begin{figure}
\centering
\includegraphics[width=\columnwidth]{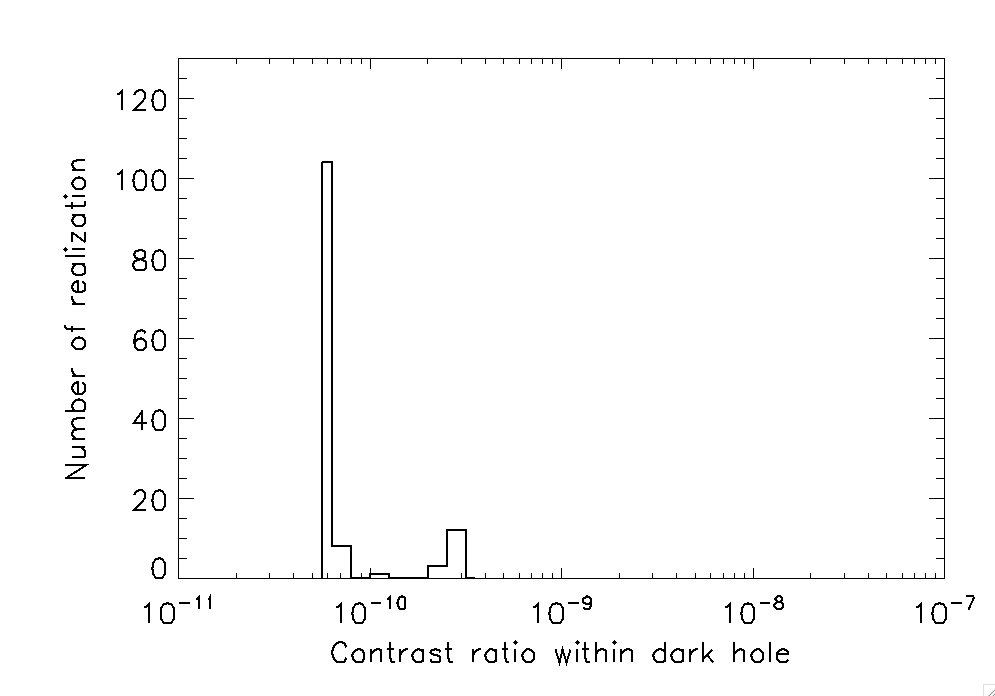}
\caption{Nominal \mbox{5$\sigma$} contrast ratio histogram within the dark hole (from 0.8 to 4 $\lambda$/D). The simulation assumes the SPEED setup at 1.65$\micron$, a perfectly co-phased mirror, a theoretical PIAACMC coronagraph and 2 DMs with infinite stroke and located at 0.2 and 1.5 m from the pupil plane.}
\label{fig:nom}
\end{figure}

The nominal case assumes \mbox{5 nm rms} aberration with a PSD in $f^{-3}$ for each passive optic including the dichro\"{i}c and the DMs windows. These parameters are used in the rest of the paper unless specified otherwise. For illustration purpose, figure \ref{fig:res_ill} shows an example of the achieved contrast ratio images: initial (top left) and coronagraphic before (top right) and after (bottom) the dark hole algorithm.
The plot on \mbox{figure \ref{fig:nom}} shows the contrast ratio histogram, representing the number of realisations (ordinate) that reaches a given \mbox{5$\sigma$} contrast (abscissa, in logarithm scale) defined as the median of the contrast computed within a dark hole from 0.8 to \mbox{4 $\lambda$/D} from the optical axis. 

Because our simulations assume a perfect wavefront sensor and preclude amplitude and temporal errors, the algorithm reaches very high contrast (nominal \mbox{5$\sigma$} contrast ratio of \mbox{$6.10^{-11}$}), well below what real instruments can achieve. Nonetheless, this nominal contrast ratio serves as a reference, showing that some of the optical parameters not appropriately set can degrade the contrast level to limiting value. 

\subsection{Phase errors}
\subsubsection{Analytical description and interpretation}
\label{sec:analytic}
The correction of an aberrated optic by a dual-DMs system depends on both the dark hole spatial frequencies and the optic location to the DMs position. 
If the DMs correction frequencies and the aberrated optic frequencies distribution are not adapted, the correction will be inefficient. 
In this section, we determine the limitation in correction 
induced by an aberrated optic at a given location. 

Following the same rationale and formalism as for \citealt{BeaulieuAbeMartinezEtAl2017}, we briefly recall here the main equations. 
An out-of-pupil optic creates at the focal plane a sine and cosine modulation as
\begin{multline}
E_f(x',y')  =  \frac{e^{i\frac{2\pi}{\lambda}(2f-z)}}{i \lambda f} \times \eta \times \ \widehat{E_{\scriptscriptstyle{opt}}(u,v)}, \\
 \label{eq:opt}
\end{multline}
with 
\begin{equation}
\mathrm{
\mbox{$\eta=cos(\frac{\pi \lambda z}{D^2}(x'^2+y'^2)) + i \ sin(\frac{\pi \lambda z}{D^2}(x'^2+y'^2))$},}
\label{eq:modulation}
\end{equation}
and where \ $\hat{}$ \  represents the Fourier Transform, $E_f$ is the electric field at the focal plane,  $E_{opt}$ is the electric field in the optic plane, $u$ and $v$ are the spatial coordinates at the optic plane, $D$ is the pupil diameter, $\lambda$ is the wavelength, $f$ is the imaging camera focal length, $z$ is the distance of the optic to the pupil plane and \textit{x'} and \textit{y'} are the spatial frequencies expressed in $\frac{\lambda}{D}$. This modulation impacts the real (cosine) and imaginary (sine) part of $\eta$ and contributes to the correction efficiency. 

We first consider the case where the optic is a DM ($\eta=\eta_{\scriptscriptstyle{DM}}$). The modulation then impacts the stroke amplitude as low sine or cosine values need to be compensated by a large DM stroke; which could be out of the algorithm linear regime assumption for energy minimisation, and even out of the DM correction range. 
\begin{figure}
\centering
\includegraphics[width=.9\columnwidth]{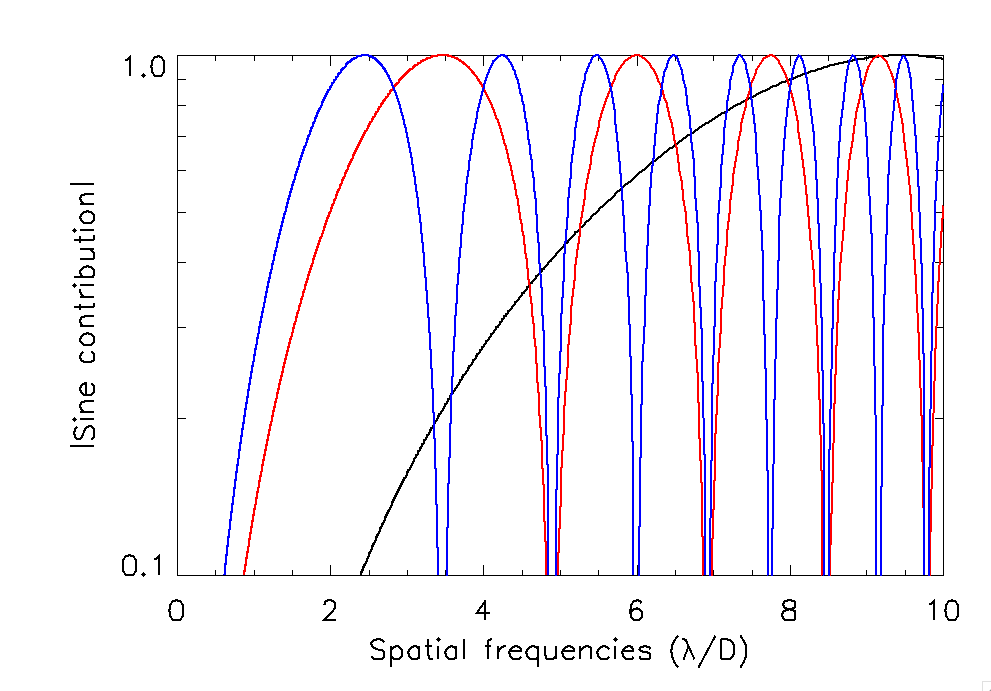}
\caption{Sine contribution as a function of the dark hole frequencies for different DM locations: 0.2 (black), 1.5 (red) and 3 m (blue) from the pupil plane.}
\label{fig:dmlocation0}
\end{figure}
We illustrate this impact by showing in figure \ref{fig:dmlocation0}, the absolute values of the sine contribution for three DM locations (0.2 m in black, 1.5 m in red and 3 m in blue), for a beam diameter of \mbox{7.7 mm} and a imaging wavelength of \mbox{1.65 $\mu m$}. The sine term (or similarly the cosine term) oscillates and thus degrades the overall system efficiency over the spatial frequencies. In particular, the sine contribution at a small separations (below \mbox{4 $\lambda$/D}) is efficiently covered by a DM located at large distance (\mbox{>1 m}, blue and red curves) but not by a DM at a small distance (black curve). 
By applying the same rationale for the cosine term and by taking into account the second DM (see \citealt{BeaulieuAbeMartinezEtAl2017} for details), the DMs locations have been defined to optimise the sine and the cosine terms between 0.8 and \mbox{4 $\lambda$/D} corresponding to DMs locations of 1.5 m and \mbox{0.2 m} from the pupil plane.

\begin{figure}
\centering
\includegraphics[width=\columnwidth]{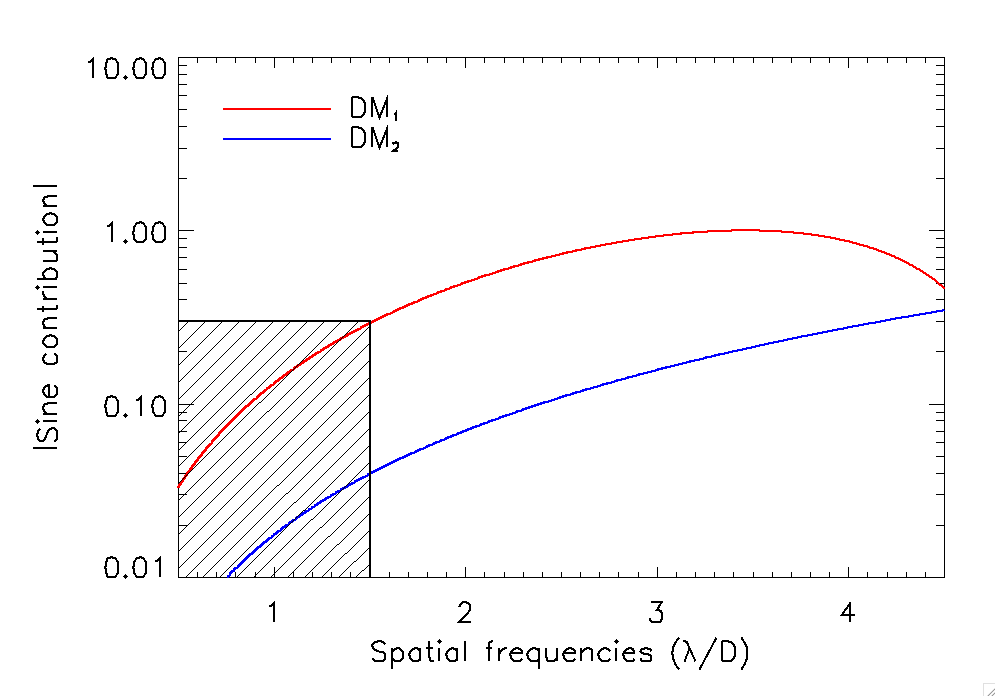}
\caption{Sine contribution of DM$_1$ (red) and DM$_2$ (blue). If the DM correction is low (dark region, sine values less than 30\%), an out-of-pupil aberrated optic will inefficiently be corrected.}
\label{fig:dm_sine}
\end{figure}

We now consider the case where the optics is an out-of-pupil aberrated window. The electric field at the focal plane is also modulated (equation \ref{eq:opt}) and needs to be corrected by the DMs. The correction efficiency depends on the DMs modulation: if the DMs modulation (sine or cosine term of $\eta_{\scriptscriptstyle{DM}}$) is low at the frequencies where the window's modulation is high, the correction will be inefficient. 
We, at first, focus on the sine contribution.
This is illustrated in figure \ref{fig:dm_sine} which shows, for the previously defined DMs (DMs locations of 1.5 and 0.2 m from the pupil plane), the 2 DMs sine contributions (red and blue curves) as a function of the spatial frequencies in \mbox{$\lambda$/D}. The dark hatched region corresponds to a poor-efficient regime, where the sine contribution of both DMs is low (arbitrarily defined as a sine value less than 30\% i.e where the DM$_{1}$ and DM$_{2}$ contributions are both lower than 30\%). 

The impact of an out-of-pupil aberrated window is determined by the contribution of this optic where the DMs are not efficient. It is the value of the window sine contribution at the frequencies where the DMs contribution is below 30\%. We thus define the degradation of correction by the mean of the window sine contribution, for frequencies where the DMs contribution is low (dark hatched region of figure \ref{fig:dm_sine}). 
This degradation term is a qualitative indicator that must be complemented by the contrast ratio estimation with end-to-end simulation.
The same rationale is applied to the cosine term of $\eta$ in equation \ref{eq:modulation}. Nevertheless, with the previously chosen DMs locations (0.2 and 1.5 m from the pupil plane), there is no region where the cosine contributions of both DMs are less than 30\%, leading to a good coverage of the real part of $\eta_{\scriptscriptstyle{DM}}$. \\
This approach is based on the DM correction degradation due to the location of an aberrated optic, i.e., the DM correction efficiency when taking into account the Talbot effect (Fresnel propagation). However, this method does not distinguishes the aberrations located before and after the coronagraph.

\subsubsection{Impact of aberrated optics in collimated beam}
\label{sec:large}
We assess the impact of aberrated optics located in a collimated beam (for instance, the dichro\"{i}c and the DMs windows), referred as to near pupil plane aberrations, even though one DM is located at 1.5 m from the pupil plane.
\begin{figure}
\centering
\includegraphics[width=\columnwidth]{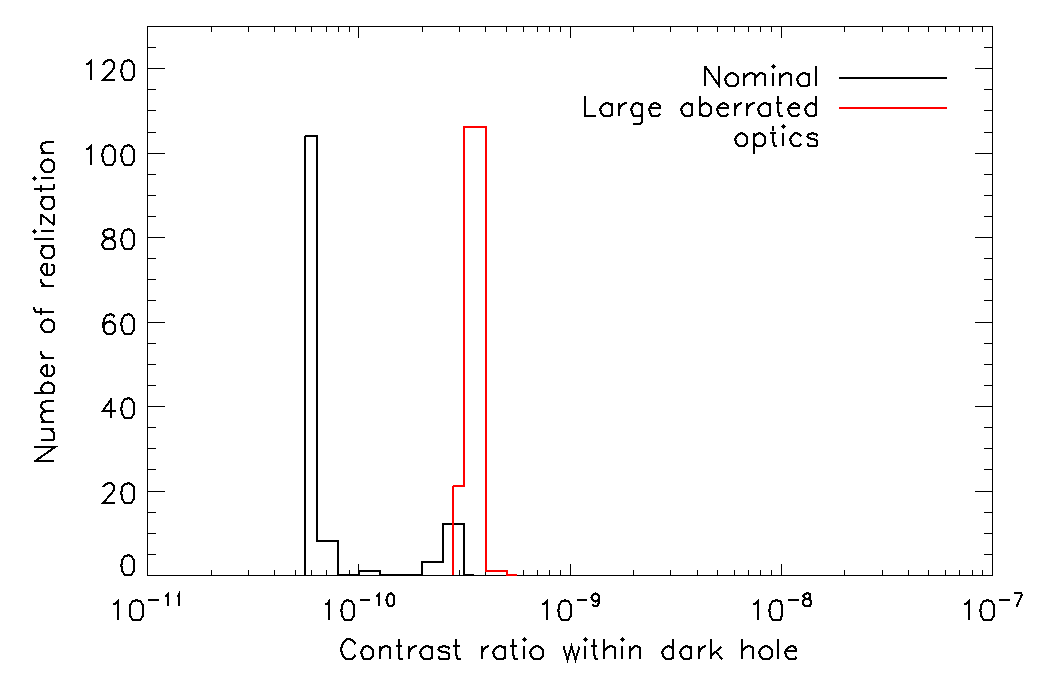}
\caption{Impact on \mbox{5$\sigma$} contrast ratio of aberrated optics located in a collimated beam.}
\label{fig:bigws}
\end{figure}
The aberration amount is based on realistic manufacturing data, i.e., \mbox{20 nm rms} for the dichro\"{i}c (standard $\lambda$/10 surface quality), \mbox{30 nm rms} for the DMs windows and \mbox{50 nm rms} for the active segmented mirror window (standard $\lambda$/4 surface quality). Each mirror window is simulated twice to take the mirror reflection into account, leading to an overall amount of aberrations of \mbox{95 nm rms}. The impact on performance is shown in \mbox{figure \ref{fig:bigws}}. Despite a large amount of aberration introduced by these optics, the contrast ratio remains high (the \mbox{5$\sigma$} contrast ratio is degraded to $3.10^{-10}$), corresponding to an efficient DMs correction. 
The DMs windows aberrations, as they are located at the DMs planes, are obviously well-corrected by the DMs themselves. 
The impact of the dichro\"{i}c is quantitatively explained by the analytical approach discussed in section \ref{sec:analytic}: by computing the degradation on correction (defined in section \ref{sec:analytic}) in this case, we find that the DMs location is appropriate to correct for this optic aberrations (degradation of correction values below 0.05).

\subsubsection{Impact of near focal-plane aberrated optics}
\label{sec:focw}
\begin{figure}
\centering
\includegraphics[width=.8\columnwidth]{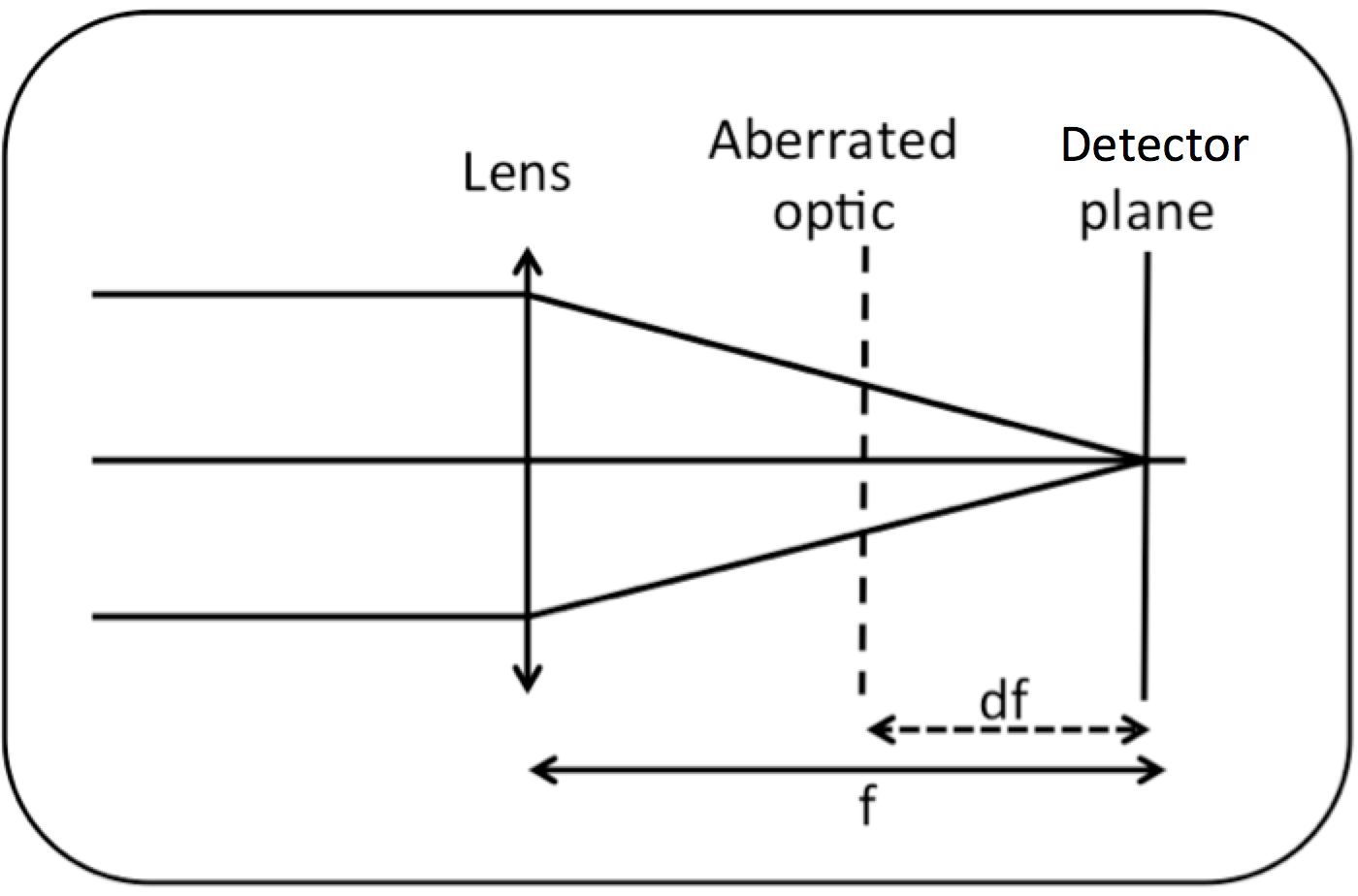}
\caption{Illustration of simulation layout: aberrated optics located at df values of 1, 2, 3, 4, 5, 10, 20, 30, 40 and 50\% of the focal length.}
\label{fig:focw_simu}
\end{figure}

\begin{figure}
\centering
\includegraphics[width=\columnwidth]{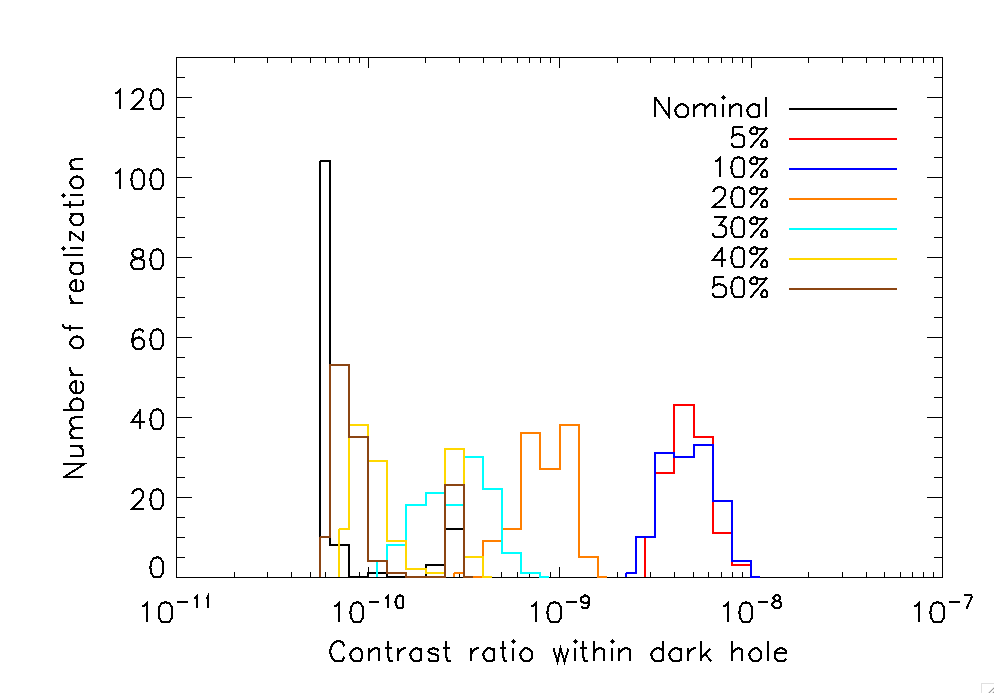}
\includegraphics[width=\columnwidth]{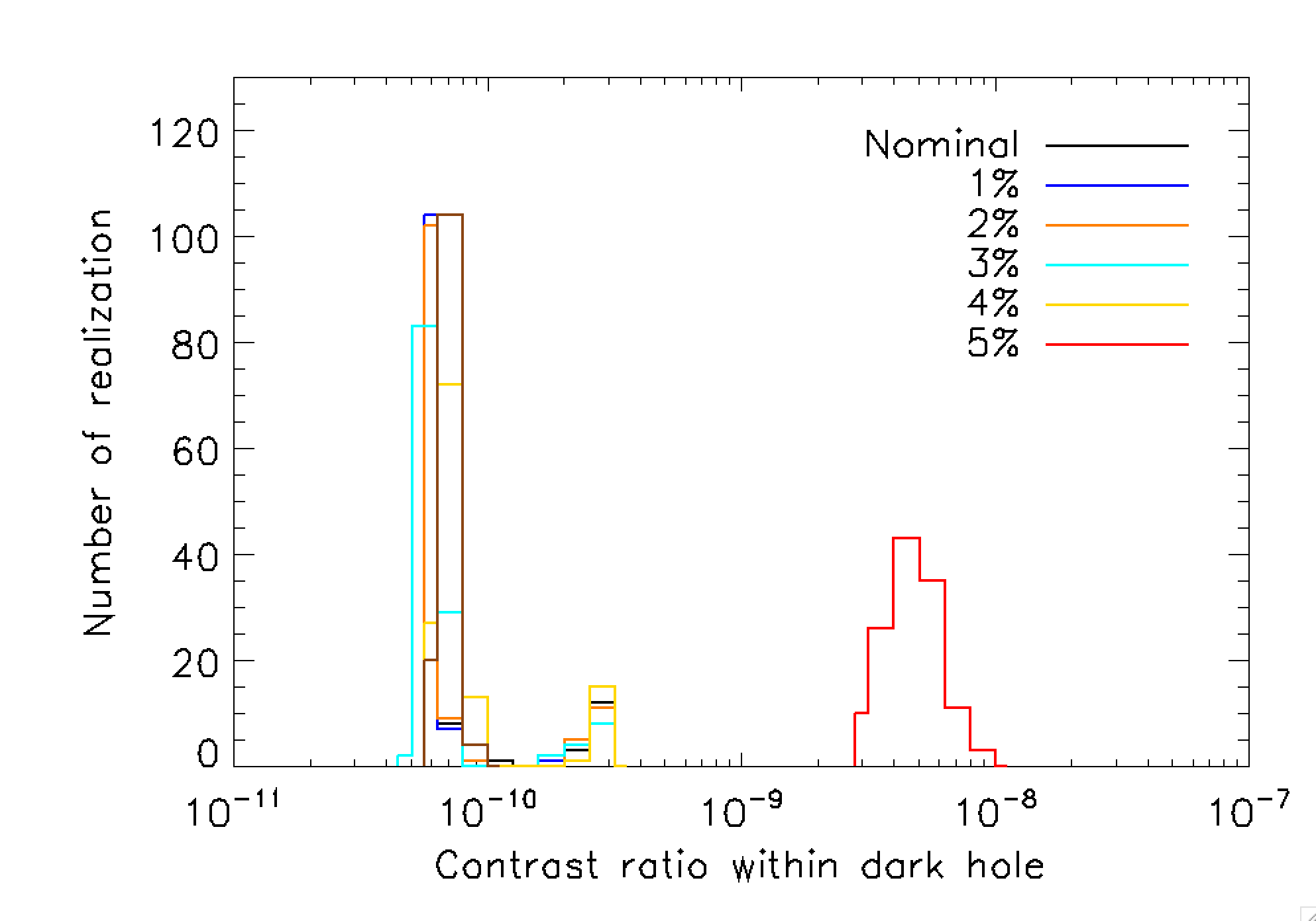}
\caption{\mbox{5$\sigma$} contrast ratio histogram for aberrated optics at df values of 5, 10, 20, 30,40 and 50\% of the focal length (top) and from 1 to 5\% of focal length (bottom).}
\label{fig:hist_focw}
\end{figure}

We here appraise the impact of a standard aberrated window (\mbox{5 nm rms}, PSD in $f^{-3}$), located near the detector plane (e.g a cryostat window). We simulate the performance when adding aberrated optics located at 1, 2, 3, 4, 5, 10, 20, 30, 40 and 50\% of the camera focal length (corresponding to $df$ in the layout of figure \ref{fig:focw_simu}).
For clarity, the results are shown on two different plots. The top plot of figure \ref{fig:hist_focw} corresponds to $df$  values of 5, 10, 20, 30,40 and 50\% of the focal length whereas the plot on the bottom corresponds to $df$ values from 1 to 5\% of the focal length. We notice a performance degradation when the aberrated optic is located near the focal plane (5 to \mbox{$\thicksim$30\%} of the focal length, see top of the figure), compared to larger locations (>30\%) but the contrast ratio is not affected for optics located very close to the focal plane (from 1 to 4\% of the focal length, see bottom of the figure). As an illustration, 5\% of the focal length corresponds to \mbox{30 mm} and 30\% corresponds to \mbox{180 mm} for the SPEED test-bed. 

\begin{figure}
\centering
\includegraphics[width=\columnwidth]{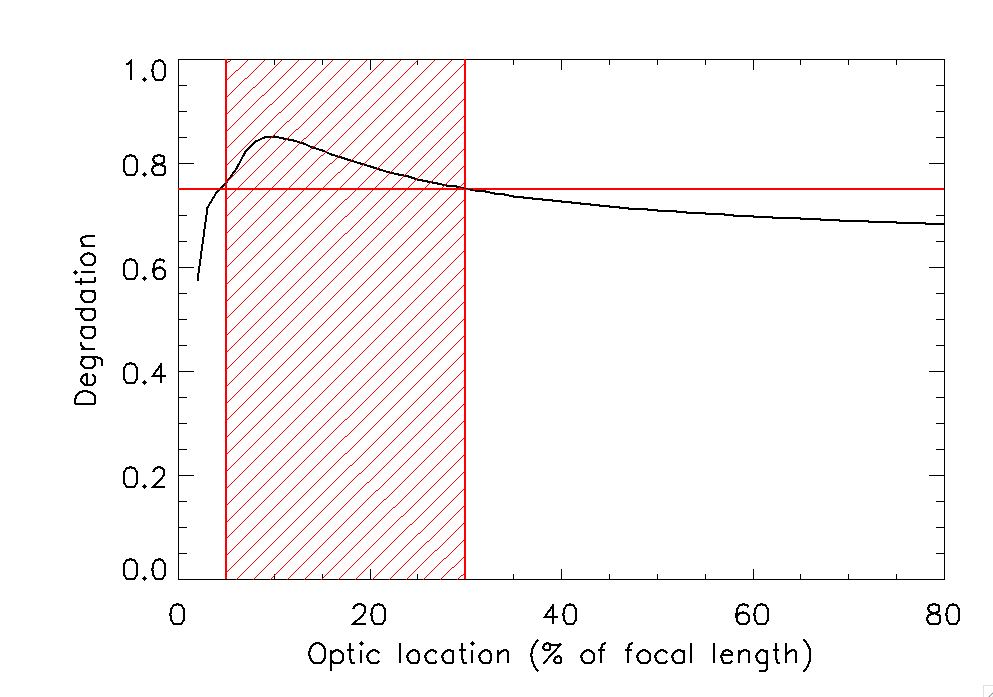}
\caption{Degradation in DMs correction due to aberrated optic, as a function of the optic location in \% of focal length.}
\label{fig:focw_deg}
\end{figure}

To grasp these results, we follow the same rationale than in section \ref{sec:analytic}. An aberrated window located near the focal plane can be considered as the image, formed by a lens, of an object located in the collimated beam. We thus compute, for each $df$ position, the geometrical location of the corresponding object in the collimated beam, and thus the corresponding distance $z$ to the pupil plane (equations \ref{eq:opt} and  \ref{eq:modulation}).
The degradation of correction defined in section \ref{sec:analytic} is finally computed for each case and is shown in figure \ref{fig:focw_deg} where we notice two contrast performance regimes. The first regime corresponds to the region where the simulated contrast ratio is significantly degraded, i.e., for $df$ values between 5 and 30\% (see figure \ref{fig:hist_focw}) where the degradation is highest (greater than 0.75, see red hatched region of figure \ref{fig:focw_deg}). The second regime is the region exhibiting limited contrast ratio degradation, where the degradation is lower than 0.75.
Thus, the previously back-of-the-envelop degradation estimation of section \ref{sec:analytic} can be used as a baseline to determine the impact in a contrast ratio of near focal plane windows or optics but end-to-end simulation is needed for quantitative estimates. We could consider in a future work implementing an estimation of these aberrations (\citealt{PaulMugnierSauvageEtAl2013}) or a non-linear approach (\citealt{PueyoNorman2013a, PaulMugnierSauvageEtAl2013}) to mitigate this effect. \\ 

Moreover, while the degradation from 40\% to 5\% is progressive (from contrast ratio of $\thicksim 10^{-10}$ to $\thicksim 5.10^{-9}$ on top of figure \ref{fig:hist_focw}), the impact for $df$ values less than 5\% is binary (contrast ratio of $5.10^{-9}$ or $\thicksim 5.10^{-11}$). Although the trend is well explained by the analytical approach of section \ref{sec:analytic}, these dichotomic values can be explained by the propagator code: \textsc{proper} \citep{Krist2007a} uses the angular spectrum
and Fresnel approximation as propagation algorithms and automatically determines which one is the best to implement from the ratio between the propagation distance and the Rayleigh distance. The transition between the two approaches is physically progressive but numerically dichotomous to avoid sampling issues. In the simulated case, the transition between the near and the far-field approach is precisely between 4 and 5\%, explaining the dichotomy in the contrast ratio values at those values. 

We also note an analogy with the effect described in \citealt{MaroisPhillionMacintosh2006} that determines the impact in polychromatic light of close to the focal plane optics in the simultaneous spectral differential imaging efficiency (SSDI, \citealt{RosenthalGurwellHo1996, Smith1987,RacineWalkerNadeauEtAl1999,MaroisDoyonRacineEtAl2000}). \citealt{MaroisPhillionMacintosh2006} evaluate how the Talbot effect due to out-of-pupil plane optics impacts the SSDI speckle-noise reduction and show that near focal-plane aberrations can significantly degrade the performance. Our study determines that, even for a monochromatic case, the effect of near focal-plane optics is not negligible on high-contrast imaging with wavefront shaping and should be taken into account during instruments development. 

These end-to-end simulations, combined with the analytical description discussed in section \ref{sec:analytic}, illustrate the impact of aberrations in the DMs correction depending on their location. Whereas the previous analysis was realized for optics near the detector plane, it can be generalized to each test-bed focal plane, upstream or downstream of the coronagraph. It is generally accepted, in the field of astronomical instrumentation, that optics aberrations upstream of the coronagraph are more detrimental than optics aberrations downstream of the coronagraph. The current results determine that, even for optics located after the coronagraph, the DMs correction can be inefficient. 

\subsubsection{Impact of phasing residuals}
\label{sec:phasing}
\begin{figure}
\centering
\includegraphics[width=\columnwidth]{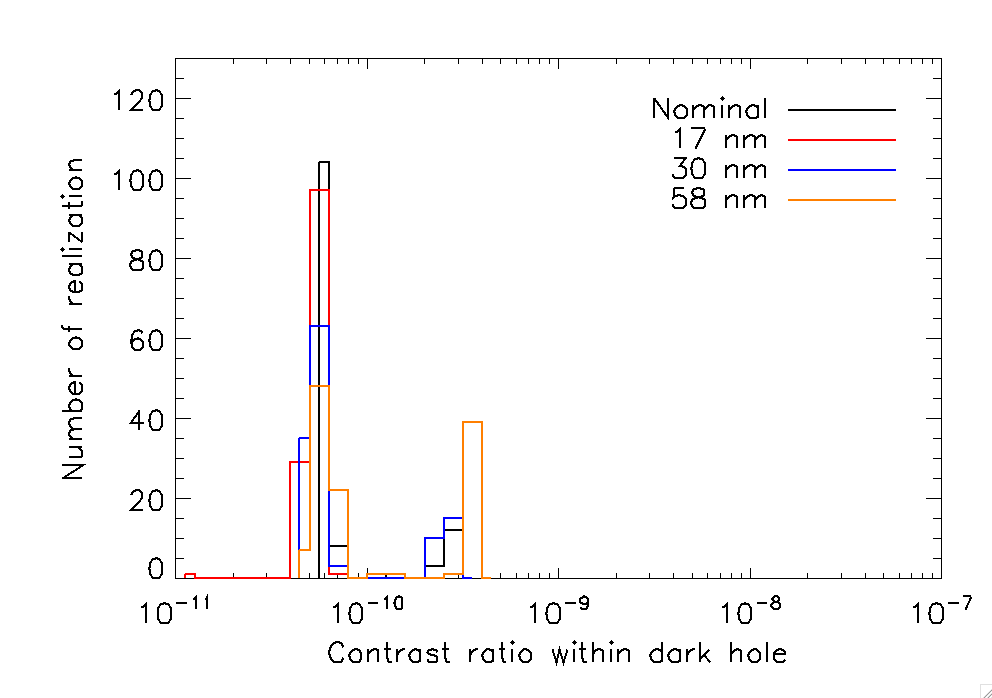}
\caption{Contrast ratio (\mbox{5$\sigma$}) histogram when taking into account 17 nm rms aberrations (10 nm defocus, 10 nm piston and 10 nm tip-tilt, in red), 30 nm rms aberrations (10 nm defocus, 20 nm piston and 20 nm tip-tilt, in blue), and 58 nm rms (10 nm defocus, 40 nm piston and 40 nm tip-tilt, in orange).}
\label{fig:cophase}
\end{figure}

In this section, we assume the presence of cophasing errors left uncorrected by the phasing unit. Because we focus on realistic laboratory test-beds, we exclude on-sky observations errors that interact with cophasing optics (i.e., low-wind effect, XAO residuals, etc.).
We simulate manufacturing and alignment errors resulting in low-order aberrations (piston, tip-tilt, and focus error on each segment). Segment defocus, that cannot be corrected by cophasing optics, is defined to \mbox{10 nm rms}, which is consistent with the SPEED ASM phase map measured in the laboratory.
We simulate the impact of cophasing errors in piston and tip-tilt through 3 cases: with 10, 20, and 40 nm rms per aberration type (piston and tip-tilt). The resulting overall aberration amounts (piston, tip-tilt, plus defocus) are respectively 17, 30 and 58 nm rms. Those values are conservative as we expect for SPEED, nearly zero residuals errors in a few cophasing algorithm iterations
(\citealt{Janin-PotironMartinezBaudozEtAl2016,Janin-PotironNDiayeMartinezEtAl2017}). As an illustration, the HiCAT test-bed achieves a residual phasing surface error of \mbox{9 nm rms} (\citealt{SoummerBradyBrooksEtAl2019}). The impact on performance (\mbox{5$\sigma$} contrast ratio) is presented in figure \ref{fig:cophase} and shows no impact on achieved contrast.  

Previous simulations point out lower tolerance in cophasing errors: the study for LUVOIR, presented in \citealt{JuanolaParramonZimmermanGroffEtAl2019}, shows that a contrast ratio of $\thicksim10^{-10}$ can be achieved with \mbox{1 nm} rms telescope aberration (piston and tip/tilt) and with wavefront control with 2 DMs. Even if it does account for the impact of more aberrated primary mirror on the performance, it seems to be a strong constraint for high-contrast achievement. This difference in high-contrast limitation can be explained by:
(i) a monochromatic analysis (the study in \citealt{JuanolaParramonZimmermanGroffEtAl2019} considers a polychromatic light), leading the algorithm to converge in a easier and more efficient way. Furthermore, the impact of aberration is reduced in the current analysis, as the simulation is in H-band (vs. in visible in \citealt{JuanolaParramonZimmermanGroffEtAl2019}) and the impact of aberration error is reduced at larger wavelengths;
(ii) the dark hole size is smaller in our analysis (overall size of \mbox{3.2$\lambda$/D} vs. \mbox{8.5$\lambda$/D} in \citealt{JuanolaParramonZimmermanGroffEtAl2019}), relaxing the strain on DMs frequencies and thus allowing more flexibility;
(iii) the LUVOIR optics distances and sizes are much larger that for the SPEED project (spatial telescope vs. laboratory test-bed), leading to larger Fresnel propagation effects.

\subsubsection{Impact of coronagraph manufacturing errors}
The PIAACMC 
consists of two aspheric mirrors to geometrically apodize the beam, a focal plane mask (FPM), and a Lyot stop. The PIAACMC components have been designed to meet the small IWA performance constraints with an ELT-like pupil (30\% central obscuration ratio, 6 spiders at \mbox{60$^{\circ}$}), in a monochromatic setup at \mbox{$1.65 \mu m$}. It has led to the definition of two centrally-symmetric mirrors for the apodization and a multi-zone phase-shifting FPM consisting in \mbox{$\thicksim$ 500} hexagons of \mbox{25 $\mu m$} diameters with depths (optical path differences) from \mbox{-0.4 to 0.4 $\mu m$}. As described in \citealt{MartinezBeaulieuGuyonEtAl2018}, those optimised optics lead to a theoretical IWA of \mbox{1.3 $\lambda$/D} and a raw contrast of $10^{-5}$ at IWA. We here simulate optical aberrations on the PIAACMC components to evaluate the impact on performance.

Several PIAACMC FPMs have been recently manufactured (i.e., \citealt{KernGuyonBelikovEtAl2016, NewmanConwayBelikovEtAl2016, KnightBrewerHamiltonEtAl2017, MartinezBeaulieuBarjotEtAl2019}). Manufactured depth errors represent typically few per cent of the peak-to-valley value of the sag. We thus simulate a FPM sag error of \mbox{5 nm rms}, consistently with the SPEED FPM manufacturing tolerance analysis (\citealt{MartinezBeaulieuGuyonEtAl2018}) and with the prototype characterisation (\citealt{MartinezBeaulieuBarjotEtAl2019}). 

\begin{figure}
\centering
\includegraphics[width=\columnwidth]{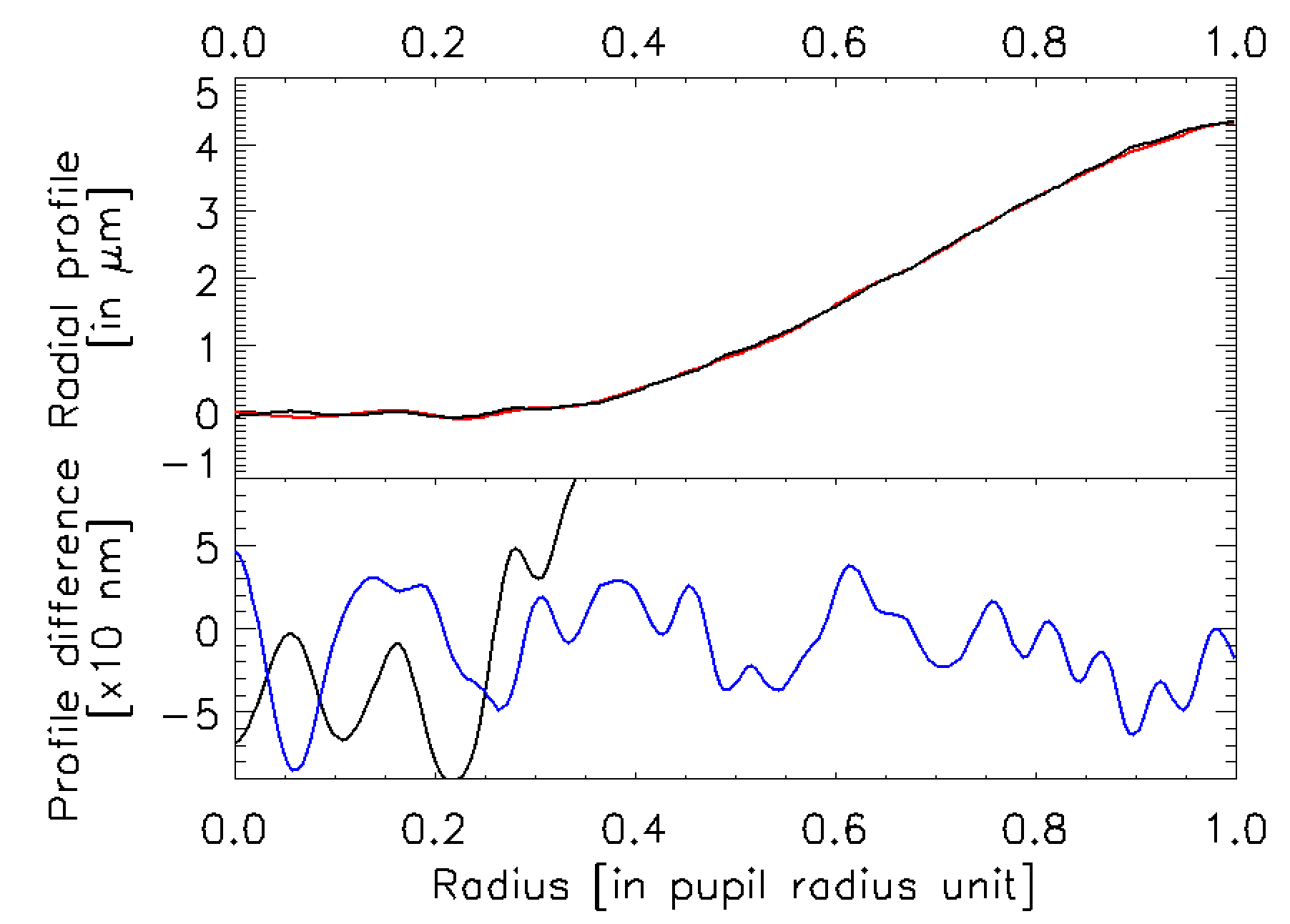}
\caption{SPEED PIAACMC first mirror profile on top: theoretical profile (dark curve on the top plot) and with 30 nm rms simulated manufacturing errors (red curve on the top plot). The  profile difference between the theory and simulated manufacturing errors is in blue in the bottom plot whereas the theoretical profile is over-plotted in dark.} 
\label{fig:m1p}
\end{figure}

We also simulate aberrations on the PIAACMC aspheric mirrors, following the defined specification: a sag deviation (best effort) of \mbox{25 nm rms} (\citealt{MartinezBeaulieuGuyonEtAl2018}). We thus simulate two cases corresponding to sag deviations of 20 and 30 nm rms, divided into low and middle frequencies, to keep smoothed mirrors profiles. The top of figure \ref{fig:m1p} illustrates the simulated SPEED PIAACMC first mirror profile  (the theoretical one in dark and with simulated manufacturing errors of 30 nm rms in red). The difference between the two profiles corresponds to the blue curve on the bottom plot of figure \ref{fig:m1p} (the theoretical profile is over-plotted in dark) . 

\begin{figure}
\centering
\includegraphics[width=\columnwidth]{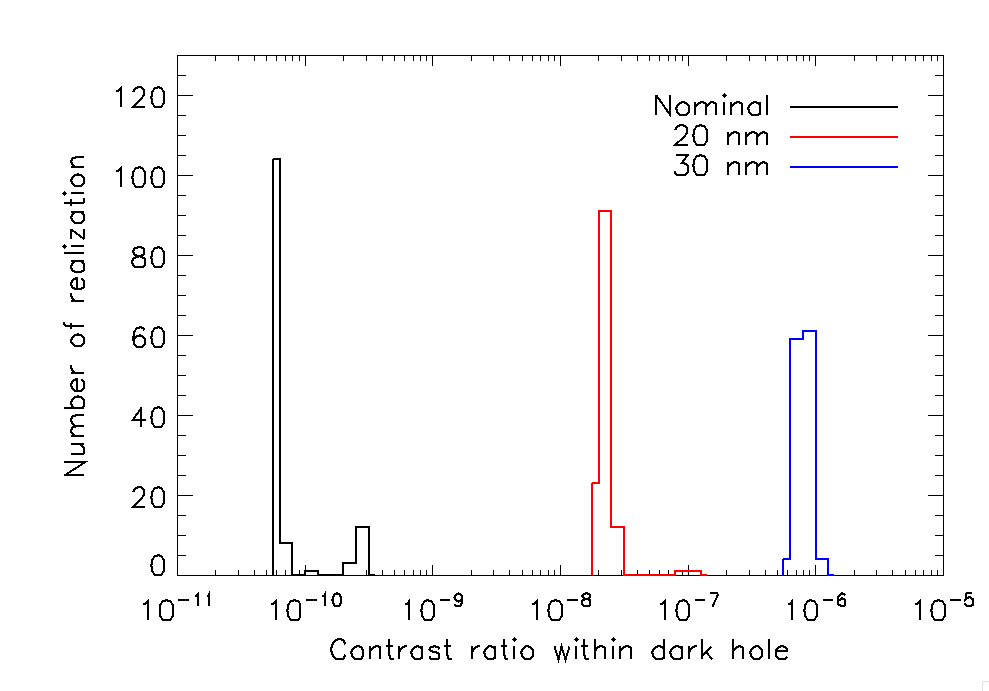}
\caption{\mbox{5$\sigma$} contrast ratio histogram assuming the simulated FPM and aspheric aberrations. The FPM is simulated with 5 nm rms aberration and the aspheric mirrors are simulated with 20 (red) and 30 (blue) nm rms.}
\label{fig:cor}
\end{figure}

The FPM and aspheric manufacturing errors have been simulated separately, showing very low impact when adding the FPM manufacturing errors and a major contribution of the aspheric mirrors errors. Figure \ref{fig:cor} represents the contrast ratio resulting in a FPM sag deviation of 5 nm rms plus aspheric mirrors sag deviations of 20 nm rms (in red) and 30 nm rms (in blue), showing the major impact of aspheric errors.

Although the aspheric mirrors are located in a collimated beam, we note a major degradation in performance for 30 nm rms errors; in contradiction with discussion in section \ref{sec:analytic} to \ref{sec:focw}. This can be explained by the contrast degradation due to coronagraphic manufacturing errors: whereas the FPM manufacturing errors are located at the focal plane and impact the overall sag, the aspheric mirrors manufacturing errors can impact the profile at high-frequencies. The theoretical profile on top of figure \ref{fig:m1p} shows high-frequencies ripples located at radii less than 0.4 pupil radius. Manufacturing errors being of the same order of magnitude as the ripples (see bottom of figure \ref{fig:m1p}) will directly degrade the coronagraph performance. The DMs will not be able to correct for this effect as long as the ripples frequencies are higher than the DM cut-off frequency and/or the DMs locations are not optimized to correct for those frequencies.

\subsection{Amplitude errors}

\subsubsection{Impact of non-uniform pupil illumination}
 The source module has been simulated using a realistic polychromatic optical fibre numerical aperture (N.A) of \mbox{$\thicksim$ 0.2} in $H$-band. This N.A range of value creates non-uniform pupil unless the focal length of the relay optic is sufficiently large to intercept only the "top-hat" of the fibre Gaussian profile, but at the price of creating Fresnel pattern due to large propagated distance. We thus simulate separately the impact of a non-uniform pupil profile coming from these two cases: (i) a non-uniform Gaussian profile due to the fibre N.A, and (ii) a non-uniform Fresnel pattern due to large propagation distances. 

\begin{figure}
\centering
\includegraphics[width=\columnwidth]{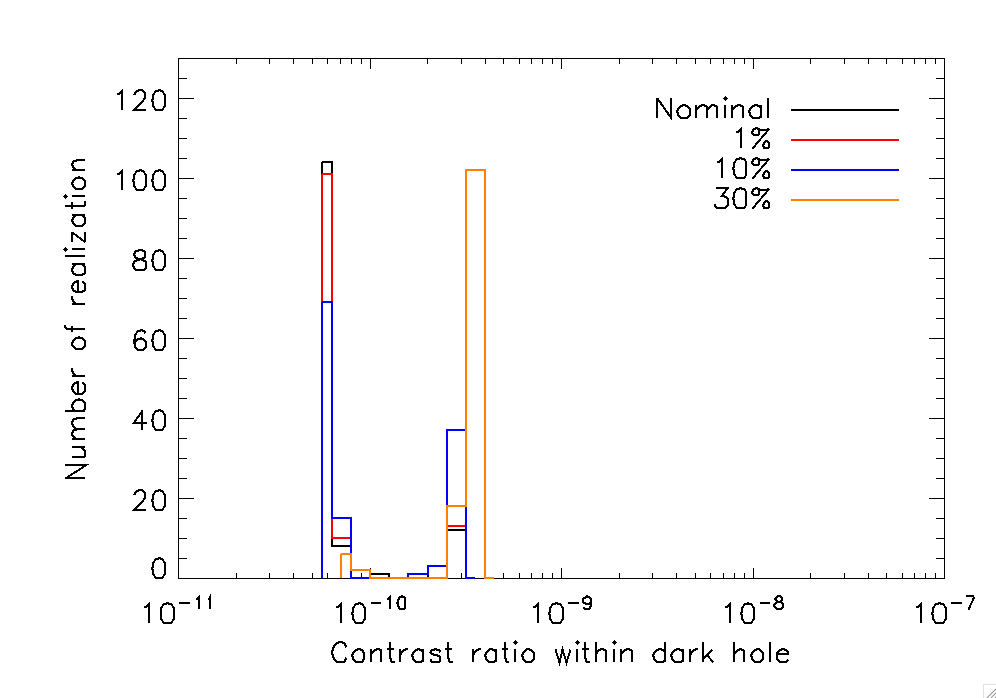}
\includegraphics[width=\columnwidth]{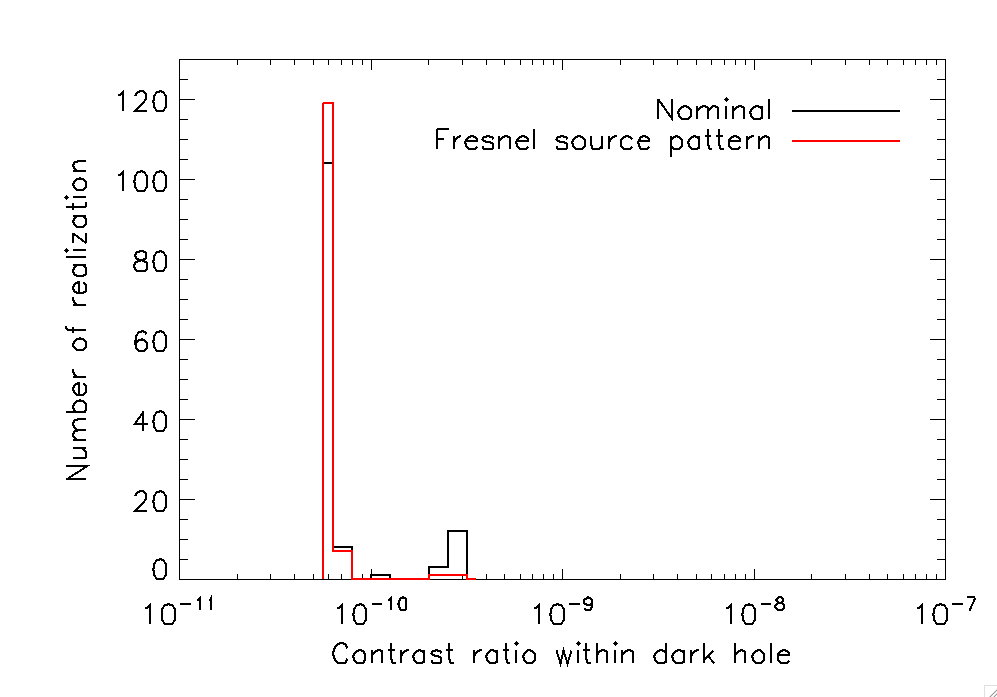}
\caption{Impact of a Gaussian non-uniform (1, 10 and 30\%) pupil (top) and of a Fresnel non-uniform pupil (bottom) on performance (\mbox{5$\sigma$} contrast ratio).}
\label{fig:source}
\end{figure}

The fibre Gaussian profile is simulated assuming 1, 10 and 30\% non-uniformity in amplitude on the overall pupil (from the top to the centre of the pupil profile). The impact of such non-uniformity is shown on the top of figure \ref{fig:source}: a non-uniformity lower than 10\% does not affect the performance as it is corrected by the two DMs. 

The Fresnel propagation pattern is simulated assuming the same optical fibre N.A (\mbox{$\thicksim$ 0.2} in $H$-band) and a focal length sufficiently large (\mbox{$\thicksim$ 600 mm}) to create a quasi-uniform (few per cent) profile ("top-hat" of the Gaussian) on the overall pupil. The impact on performance (\mbox{5$\sigma$} histogram) is shown on the bottom of figure \ref{fig:source}. We do not see any impact of such profile on the contrast ratio performance.

We can observe that the amplitude errors from a non-uniform pupil illumination are well corrected by the system. The wavefront shaping is efficient because of the dual-DM architecture (that correct for both phase and amplitude errors) and the amplitude error localisation (near the pupil plane).


\subsubsection{Impact of ASM missing segments}
\begin{figure}
\centering
\includegraphics[width=.85\columnwidth]{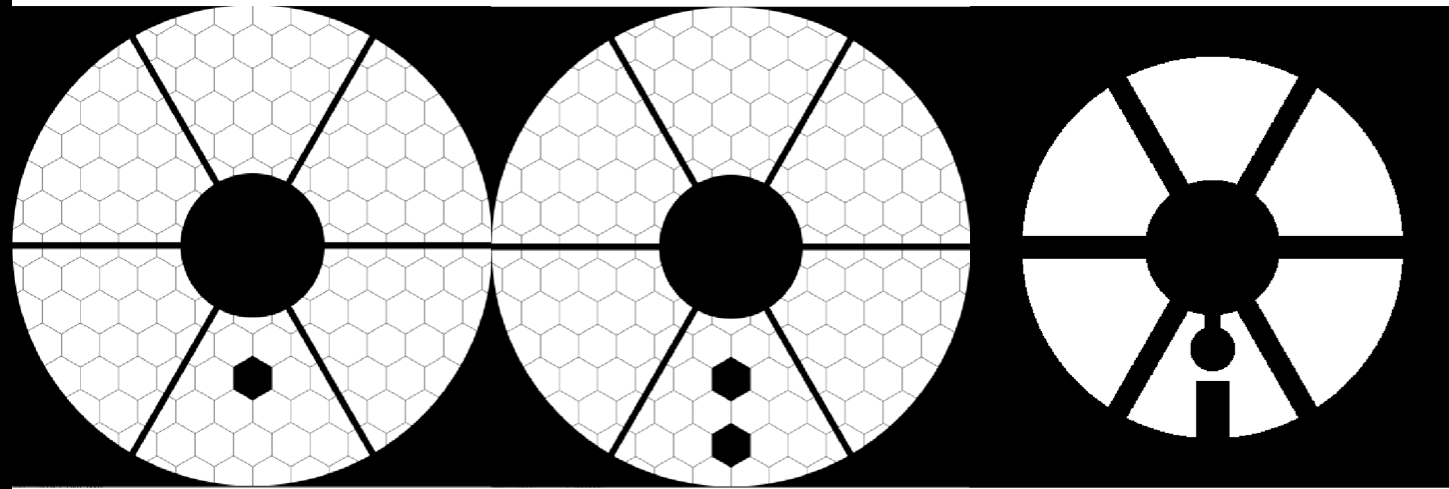}
\includegraphics[width=\columnwidth]{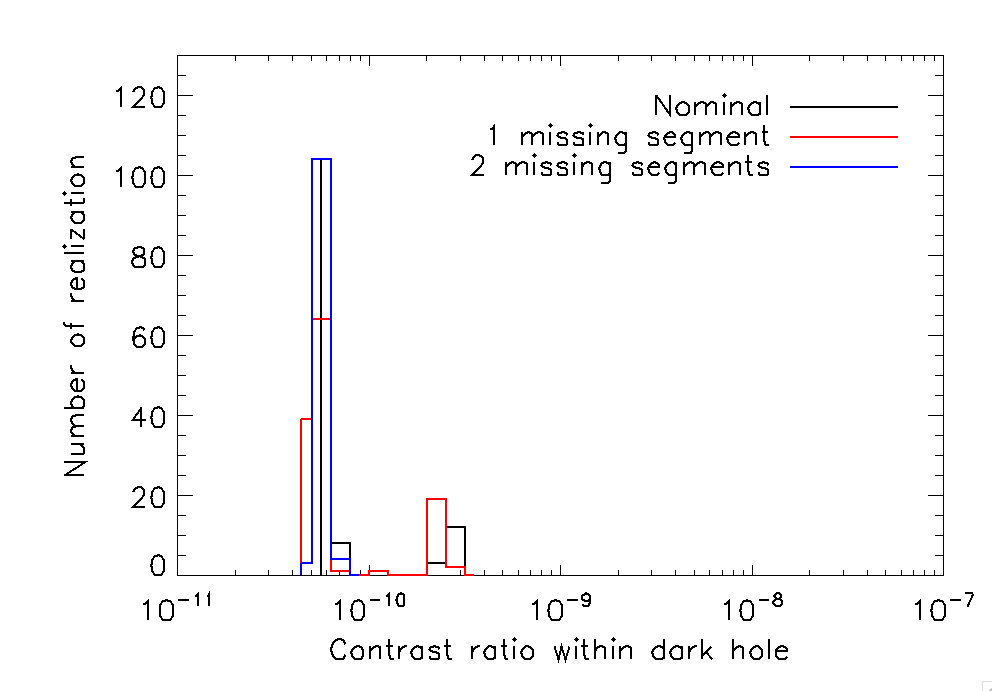}
\caption{Pupil with 1 (top left) and 2 (top centre) missing segments and adapted Lyot stop for the 2 missing segments cases (top right). The impact on performance is shown on the bottom plot.}
\label{fig:missingseg}
\end{figure}

In this section, we investigate the impact of a few ASM non-functional segments on the performance. We choose to study one of the worst case, where the missing segments are not obstructed by the spiders, for the sake of generality.  
Figure \ref{fig:missingseg} shows the results for one (top left) and two missing segments (top centre). The Lyot stop is designed to cover the missing segments for each case (e.g., top right figure). We observe no impact on the \mbox{5$\sigma$} contrast ratio (bottom plot).

No major impact on performance is observed when increasing the number of missing segments while adapting the Lyot stop. 
Even though we see no significant impact, ASM missing segments can impact the performance, by degrading the cophasing correction level and modifying the overall instrument transmission and signal-to-noise ratio (SNR). 


\subsubsection{Impact of ASM reflectivity variation}
\begin{figure}
\centering
\includegraphics[width=\columnwidth]{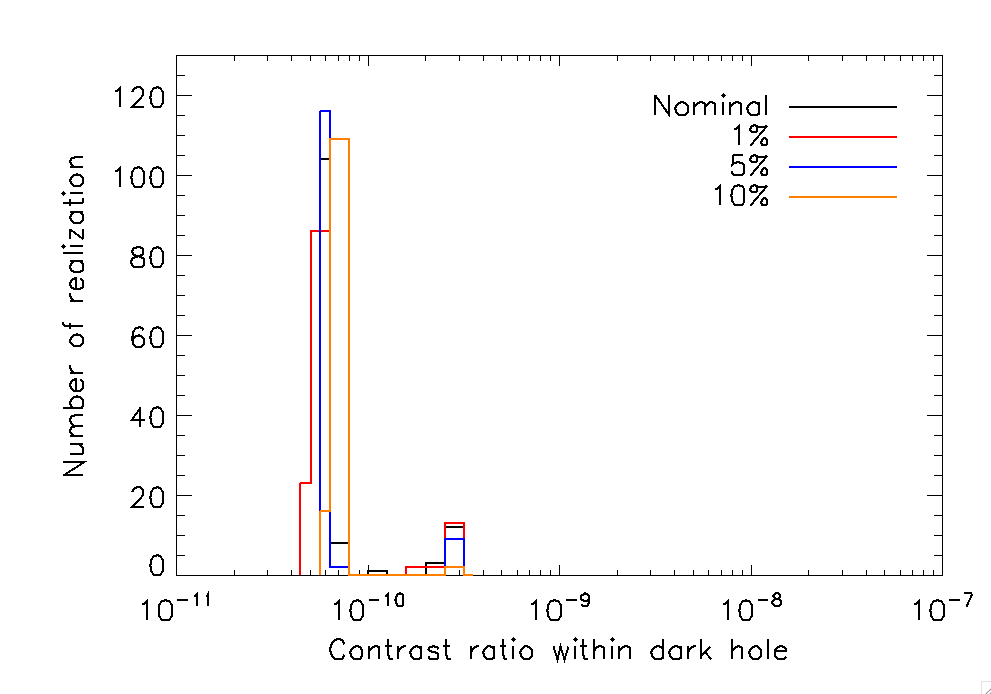}
\caption{\mbox{5$\sigma$} contrast ratio histogram for the nominal case (black) and for segment reflectivity variations of 1, 5, and 10\% in red, blue, and orange, respectively.}
\label{fig:refl}
\end{figure}

We consider the effect of segment reflectivity variations by analysing the impact of three configurations:
(i) a mean reflectivity of 99\% over the pupil with a variation of $\pm$1\%, (ii) a mean reflectivity of 95\% with a variation of $\pm$5\%, and (iii) a mean reflectivity of 90\% with a variation of $\pm$10\%. 
Results are shown on figure \ref{fig:refl}, where reflectivity variations less than 10\% does not significantly impact the contrast, illustrating that the locations of the two DMs lead to efficient errors correction at the pupil plane (see section \ref{sec:large}). 
While missing segments and segment reflectivity variations create speckles in the coronagraphic image, these static errors are well corrected by the active optic because we use a dual-DMs architecture and because the errors are localised at the pupil plane, where the wavefront shaping is efficient.

\subsection{Errors from the active correction system}

\subsubsection{Impact of deformable mirror finite stroke}
We assess the impact of DMs stroke precision in the dark hole algorithm efficiency. The algorithm was adapted to consider the smallest step an actuator can achieve. This DM stroke precision is evaluated from the manufacturer data. The deflection curve (deflection vs. voltage) and the driver precision (14 bits) lead to a maximum stroke precision of \mbox{0.25 nm} with an average of \mbox{0.15 nm}. We thus study the impact of representative stroke precision values (of 0.1, 0.2 and \mbox{0.5 nm}) on the contrast performance. For simplicity, we assume in the rest of the analysis, that this stroke precision value is the same for each actuator of the 2 DMs. 

\begin{figure}
\centering
\includegraphics[width=\columnwidth]{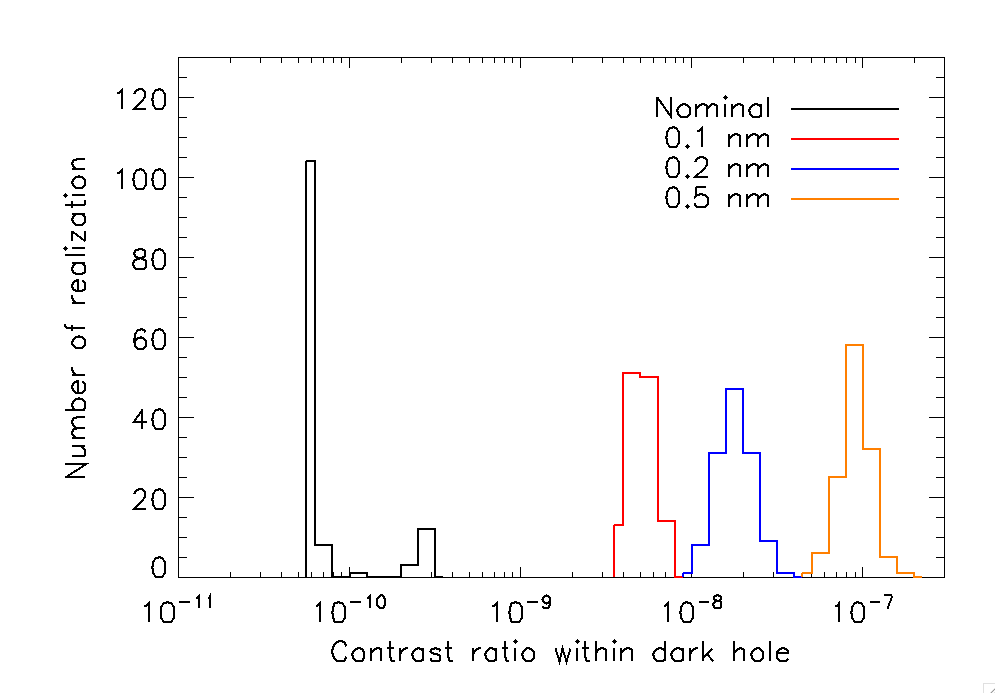}
\caption{Contrast ratio (\mbox{5$\sigma$}) histogram when taking into account 0.1, 0.2, and 0.5 nm finite stroke.}
\label{fig:stroke}
\end{figure}

 The degradation in contrast ratio due to finite stroke was described in \citealt{JiangLingRaoEtAl1991} and \citealt{TraugerMoodyGordonEtAl2011}. \citealt{TraugerMoodyGordonEtAl2011} defines, as a rule of thumb, the contrast floor due to DM error fitting error by $\thicksim \pi \frac{(8\sigma)^2}{(n\lambda)^2}$. This value can be computed by assuming a wavefront corrected at $n$ locations ($n$ defined as the DM actuators number) over the aperture diameter with a stroke precision defined as $\sigma_{0}$. For simplicity, we assume a DM located at the pupil plane. The fitting error variance over the whole aperture (defined as $\mathscr{A}=\pi n^2/4$) is thus $\sigma_{\scriptscriptstyle{err}}=4\sigma_0^2/(\pi n^2)$. The corresponding electric field at the focal plane is
\begin{equation}
E_f \approx \widehat{A(u,v)}+\widehat{iA(u,v)\phi(u,v)}  \approx \widehat{iA(u,v)\phi(u,v)},
\label{eq:var}
\end{equation}
when assuming small phase and a setup with a coronagraph that removes (or significantly attenuates) the constant term $A$. The phase $\phi$ is defined as \mbox{$2\times \frac{2\pi}{\lambda} \sigma_{\scriptscriptstyle{err}}$} assuming a reflective (mirror) surface.
We can write the contrast inside the dark hole using the Parseval theorem by 
\begin{equation}
\mathscr{C}_{\scriptscriptstyle{DH}} \approx (2\times \frac{2\pi}{\lambda})^2 \sigma_{\scriptscriptstyle{err}}^2 \approx \pi \frac{(8\sigma_0)^2}{(n\lambda)^2},
\label{eq:stroke}
\end{equation}
in accordance with \citealt{TraugerMoodyGordonEtAl2011}. Equation \ref{eq:stroke} leads to contrast floor values of $6.10^{-10}$, $2.10^{-9}$, and $2.10^{-8}$ for stroke precision values of 0.1, 0.2, and 0.5 nm, respectively. 
The \mbox{5$\sigma$} contrast ratio histograms for stroke precision values of 0.1, 0.2, and 0.5 nm are shown in \mbox{figure \ref{fig:stroke}}. 
As illustrated in table \ref{tab:stroke}, the means of the achieved \mbox{5$\sigma$} contrasts are $5.10^{-9}$, $2.10^{-8}$, and $9.10^{-8}$, and correspond to median contrast values of $7.10^{-10}$, $2.10^{-9}$, and $1.10^{-8}$, consistent with the theoretical values of $6.10^{-10}$, $2.10^{-9}$ and $2.10^{-8}$.
Whereas equation \ref{eq:stroke} assumes a single DM configuration at the pupil plane, it fairly estimates the contrast limitation due to finite stroke in the case of two out-of-pupil plane DMs.

\subsubsection{Impact of deformable mirror non-functional actuators}
\begin{figure}
\centering
\includegraphics[width=\columnwidth]{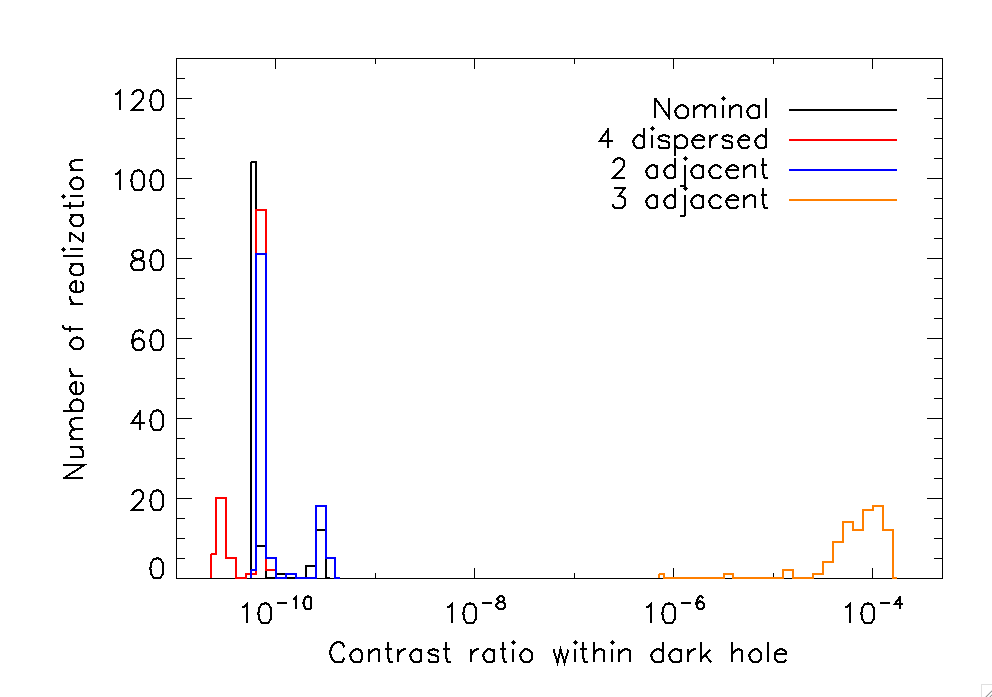}
\caption{Contrast ratio (\mbox{5$\sigma$}) histogram when taking into account 1 (red curve), 2 (blue curve) non-functional actuator on one DM or 1 non-functional actuator per DM (orange curve).}
\label{fig:act}
\end{figure}

We appraise the impact of non-functional actuators on the deformable mirrors for the wavefront shaping, as illustrated in figure \ref{fig:act}. We simulate the impact in two cases: when the dead actuators are dispersed over the DM (dispersed actuators, red curve) or adjacent (two or three grouped actuators for the blue and orange curves). 
The deformable mirrors are located out-of-the pupil plane, at a large distance from the pupil plane, such that the image of a dead actuator is \textit{diluted} at the Lyot stop plane, precluding the use of an adapted Lyot stop to prevent this effect. This also explains that the performance is less affected by dead actuators scattered over the DM than adjacent actuators (where the impact is less \textit{diluted}).

\citealt{MatthewsCreppVasishtEtAl2017} analysed the impact of DM damaged actuators on the simulated performance of Project 1640. The authors simulated some damaged actuators on a DM located at the pupil plane and showed low contrast degradation for randomly distributed and dispersed dead actuators (zero stroke value). Our analysis shows the same low degradation for dispersed actuators but also enlightens the impact of grouped dead actuators on performance. 
\begin{table}
\begin{centering}
\begin{tabular}{| m{0.2\columnwidth} | m{0.2\columnwidth} | m{0.2\columnwidth} | m{0.2\columnwidth} |}
\hline
\begin{center} Stroke precision  \end{center} &  \begin{center} Theoretical mean contrast \end{center} & \begin{center} Achieved mean contrast \end{center} & \begin{center} Achieved $5\sigma$ contrast \end{center} \\ 
\hline
\begin{center} 0.1 nm \end{center} & \begin{center} $6.10^{-10}$ \end{center} & \begin{center} $7.10^{-10}$ \end{center} & \begin{center} $5.10^{-9}$  \end{center} \\ 
\hline
\begin{center} 0.2 nm \end{center}& \begin{center} $2.10^{-9}$ \end{center} & \begin{center} $2.10^{-9}$ \end{center} & \begin{center} $2.10^{-8}$ \end{center} \\ 
\hline
\begin{center} 0.5 nm \end{center} & \begin{center} $2.10^{-8}$ \end{center} & \begin{center} $1.10^{-8}$ \end{center} & \begin{center} $9.10^{-8}$ \end{center} \\ 
\hline
\end{tabular}
\end{centering}
\caption{Theoretical and achieved contrast ratio for stroke precision of 0.1, 0.2, and 0.5 nm. The theoretical mean contrast is computed from equation \ref{eq:stroke} at 1.65$\micron$ with n=34 actuators.}
\label{tab:stroke}
\end{table}


We also simulated this impact with a larger dark hole (from 2 to \mbox{5 $\lambda$/D}). It showed the same degradation trend, illustrating that this impact is not inherent to the small separation high-contrast paradigm.

\section{Conclusion and discussion}
In \citealt{BeaulieuAbeMartinezEtAl2017}, we demonstrated that imaging at small angular separations requires a large setup and a small dark hole size. The analysis only considered the wavefront shaping system parameters such as the number of actuators, the deformable mirror locations, and the optic aberrations (level and frequency distribution).
In that context, we used an ideal and generic high-contrast architecture with a perfect coronagraph, a monolithic circular aperture without any central obscuration nor spiders, etc.
In the present study, we have extended the former study to a more realistic setup, combining a segmented and obstructed telescope pupil with a real coronagraph (PIAACMC). We also have included: (i) phase aberrations such as aberrated optics located in a collimated beam (e.g., DM windows or dichro\"{i}c) or near the focal plane (e.g., camera cryostat window), telescope phasing residuals or coronagraph manufacturing errors; (ii) amplitude errors such as non-uniform pupil illumination, primary mirror missing segments and segment reflectivity variations; (iii) errors originating from the wavefront shaping system itself (finite stroke and non-functional actuators). We have simulated the impact of these realistic parameters on contrast ratio at very small separations (around \mbox{1 $\lambda$/D}). 

We summarise the impact of each item on the performance in \mbox{Table \ref{tab:resume}}, where the most limiting parameters are written in bold for the sake of clarity.
We notice that a highly aberrated optic does not impact the performance as long as it is located in a collimated beam (cophasing errors, deformable mirrors windows, dichroïc). We also found that one or two missing segments can be compensated by adapting the Lyot stop accordingly. 
We have identified some major constraints to high contrast at small separations that, to our knowledge, were not identified in previous analysis. Some of these constraints comes from the corrected system itself (coronagraph and deformable mirror). For instance, finite DM stroke, adjacent non-functional actuators, as well as phase errors on the coronagraph or near the focal plane are a major limitation to high-contrast imaging at small separations. They significantly degrade the contrast ratio (arbitrarily set to a degradation of \mbox{~1 000}). More specifically, in the astronomical instrumentation community, a clear dichotomy is widely accepted between optics aberrations upstream and downstream of the coronagraph. Our study clearly demonstrates that optics aberrations downstream of the coronagraph cannot be neglected. It will severely impact the wavefront shaping performance otherwise.


We demonstrate that the contrast limit due to the DM stroke as predicated by equation \ref{eq:stroke} (\citealt{TraugerMoodyGordonEtAl2011}) is fairly valid. We propose a methodology (see section \ref{sec:analytic}) to estimate the correction efficiency in the presence of the aberrated window location, regardless of its position on the optical design. 
We note that this conclusion is valid for the  SPEED setup, and especially for the PIAACMC coronagraph, even if the SPEED optical model is representative to high-contrast imaging test-beds.
An end-to-end simulation of high-contrast setup is always needed to assess the expected contrast value. Nonetheless, the parameters studied in this analysis can be used as a baseline to assess some major limitations for high-contrast imaging at small separation, for ground or space-based instrumentation or for other test-beds. We demonstrate that achieving high-contrast below \mbox{2 $\lambda$/D} is challenging and needs an appropriate setup.

In future work, we will continue to improve the realism of our end-to-end modelling by adding realistic wavefront sensor and detector noises. We also aim at testing (i) the non-linear dark hole approach (\citealt{PueyoNorman2013a, PaulMugnierSauvageEtAl2013}) and in particular if it can compensate for non-functional actuators and (ii) broadband performance.
More peculiar and recent sources of errors could also be envisioned as well such as the petal phase offset (\citealt{NDiayeMartinacheJovanovicEtAl2018}) and the low wind effect (\citealt{SauvageFuscoLambEtAl2016}).
Because of the intrinsic faintness of low mass exoplanets, integration time will be on the order of hours, stability of the dark-hole over these timescales is critical, and quasi-static speckles will need to be treated as well.

\begin{table}
\begin{centering}
\begin{tabular}{| m{0.7\linewidth} | >{\centering\arraybackslash}m{0.25\linewidth} |}
\hline
\begin{center} \textsc{\textbf{Parameter}}  \end{center} &  \begin{center} $5\sigma$ \textsc{\textbf{contrast ratio}} \end{center} \\ 
\hline
Nominal & 6.$10^{-11}$  \\ 
\hline
 \multicolumn{2}{|c|}{\textbf{Phase errors}} \\
\hline
Aberrated windows in collimated beam &  3.$10^{-10}$ \\ 
\hline
\textbf{Aberrated windows near the science detector \newline (5 to 30\% of focal length)} &  $\mathbf{>10^{-9}}$ \\ 
\hline
Aberrated windows near the science detector \newline (1 to 4\% or 30\% of focal length) &  <$10^{-9}$ \\ 
\hline
Phasing residual wavefront error &  <5.$10^{-10}$  \\ 
\hline
Realistic PIAACMC coronagraph with 20 nm rms on aspherics mirrors & 2.$10^{-8}$ \\ 
\hline
\textbf{Realistic PIAACMC coronagraph with 30 nm rms on aspherics mirrors} & $\mathbf{8.10^{-7}}$ \\ 
\hline
\multicolumn{2}{|c|}{\textbf{Amplitude errors}} \\
\hline
Non uniform source (if <10$\%$) & 6.$10^{-11}$  \\ 
\hline
ASM non functional segments with adapted Lyot stops & 6.$10^{-11}$ \\ 
\hline
ASM reflectivity variation  &  <5.$10^{-10}$  \\ 
\hline
\multicolumn{2}{|c|}{\textbf{Correction errors}} \\
\hline
DMs stroke 0.1 nm &  $5.10^{-9}$   \\ 
\hline
DMs stroke 0.2 nm & $2.10^{-8}$   \\ 
\hline
\textbf{DMs stroke 0.5 nm} &  $\mathbf{9.10^{-8}}$   \\ 
\hline
\textbf{3 adjacent non functional actuators} &  $\mathbf{10^{-4}}$   \\ 
\hline
\end{tabular}
\end{centering}
\caption{Impact of each simulated parameter on the $5\sigma$ contrast ratio.}
\label{tab:resume}
\end{table}

\section*{Acknowledgements}
The authors are thankful to the computing centre personal for their support. 

\section*{Data Availability}
The data underlying this article will be shared on reasonable request to the corresponding author.


\bibliographystyle{mnras}
\bibliography{test0}

\begin{thebibliography}{}
\makeatletter
\relax
\def\mn@urlcharsother{\let\do\@makeother \do\$\do\&\do\#\do\^\do\_\do\%\do\~}
\def\mn@doi{\begingroup\mn@urlcharsother \@ifnextchar [ {\mn@doi@}
  {\mn@doi@[]}}
\def\mn@doi@[#1]#2{\def\@tempa{#1}\ifx\@tempa\@empty \href
  {http://dx.doi.org/#2} {doi:#2}\else \href {http://dx.doi.org/#2} {#1}\fi
  \endgroup}
\def\mn@eprint#1#2{\mn@eprint@#1:#2::\@nil}
\def\mn@eprint@arXiv#1{\href {http://arxiv.org/abs/#1} {{\tt arXiv:#1}}}
\def\mn@eprint@dblp#1{\href {http://dblp.uni-trier.de/rec/bibtex/#1.xml}
  {dblp:#1}}
\def\mn@eprint@#1:#2:#3:#4\@nil{\def\@tempa {#1}\def\@tempb {#2}\def\@tempc
  {#3}\ifx \@tempc \@empty \let \@tempc \@tempb \let \@tempb \@tempa \fi \ifx
  \@tempb \@empty \def\@tempb {arXiv}\fi \@ifundefined
  {mn@eprint@\@tempb}{\@tempb:\@tempc}{\expandafter \expandafter \csname
  mn@eprint@\@tempb\endcsname \expandafter{\@tempc}}}

\bibitem[\protect\citeauthoryear{{Baudoz}, {Boccaletti}, {Baudrand}  \&
  {Rouan}}{{Baudoz} et~al.}{2006}]{BaudozBoccalettiBaudrandEtAl2006}
{Baudoz} P.,  {Boccaletti} A.,  {Baudrand} J.,   {Rouan} D.,  2006, in {Aime}
  C.,  {Vakili} F.,  eds, IAU Colloq. 200: Direct Imaging of Exoplanets:
  Science Techniques. pp 553--558, \mn@doi{10.1017/S174392130600994X}

\bibitem[\protect\citeauthoryear{{Baudoz}, {Galicher}, {Potier}, {Dupuis},
  {Thijs}  \& {Patru}}{{Baudoz} et~al.}{2018}]{BaudozGalicherPotierEtAl2018}
{Baudoz} P.,  {Galicher} R.,  {Potier} A.,  {Dupuis} O.,  {Thijs} S.,   {Patru}
  F.,  2018, in Advances in Optical and Mechanical Technologies for Telescopes
  and Instrumentation III. p. 107062O, \mn@doi{10.1117/12.2314089}

\bibitem[\protect\citeauthoryear{{Beaulieu}}{{Beaulieu}}{2017}]{Beaulieu2017}
{Beaulieu} M.,  2017, PhD thesis, France: Universite Cote d'Azur;
  2017.~Publication Number: tel-01599054

\bibitem[\protect\citeauthoryear{{Beaulieu}, {Abe}, {Martinez}, {Baudoz},
  {Gouvret}  \& {Vakili}}{{Beaulieu}
  et~al.}{2017}]{BeaulieuAbeMartinezEtAl2017}
{Beaulieu} M.,  {Abe} L.,  {Martinez} P.,  {Baudoz} P.,  {Gouvret} C.,
  {Vakili} F.,  2017, Monthly Notices of the Royal Astronomical Society, 469,
  218

\bibitem[\protect\citeauthoryear{{Beuzit} et~al.,}{{Beuzit}
  et~al.}{2008}]{BeuzitFeldtDohlenEtAl2008}
{Beuzit} J.-L.,  et~al., 2008, in Ground-based and Airborne Instrumentation for
  Astronomy II. p. 701418, \mn@doi{10.1117/12.790120}

\bibitem[\protect\citeauthoryear{{Bolcar} et~al.,}{{Bolcar}
  et~al.}{2018}]{BolcarCrookeHylanEtAl2018}
{Bolcar} M.~R.,  et~al., 2018, in Space Telescopes and Instrumentation 2018:
  Optical, Infrared, and Millimeter Wave. p. 106980O,
  \mn@doi{10.1117/12.2313350}

\bibitem[\protect\citeauthoryear{{Bord{\'e}} \& {Traub}}{{Bord{\'e}} \&
  {Traub}}{2006}]{BordeTraub2006}
{Bord{\'e}} P.~J.,  {Traub} W.~A.,  2006, \mn@doi [\apj] {10.1086/498669},
  \href {http://adsabs.harvard.edu/abs/2006ApJ...638..488B} {638, 488}

\bibitem[\protect\citeauthoryear{Bottom, Femenia, Huby, Mawet, Dekany, Milburn
  \& Serabyn}{Bottom et~al.}{2016}]{BottomFemeniaHubyEtAl2016}
Bottom M.,  Femenia B.,  Huby E.,  Mawet D.,  Dekany R.,  Milburn J.,   Serabyn
  E.,  2016, in Adaptive Optics Systems V. p. 990955

\bibitem[\protect\citeauthoryear{{Foo}, {Palacios}  \& {Swartzlander}}{{Foo}
  et~al.}{2005}]{FooPalaciosSwartzlander2005}
{Foo} G.,  {Palacios} D.~M.,   {Swartzlander} Jr. G.~A.,  2005, \mn@doi [Optics
  Letters] {10.1364/OL.30.003308}, \href
  {http://adsabs.harvard.edu/abs/2005OptL...30.3308F} {30, 3308}

\bibitem[\protect\citeauthoryear{{Garrett}, {Crill}, {Patterson}, {Mejia},
  {Seo}  \& {Siegler}}{{Garrett} et~al.}{2019}]{GarrettCrillEtAl2019}
{Garrett} R.,  {Crill} B.,  {Patterson} K.,  {Mejia} C.,  {Seo} B.,   {Siegler}
  N.,  2019, in {{}}. \url
  {https://exoplanets.nasa.gov/internal_resources/1170/}

\bibitem[\protect\citeauthoryear{{Gaudi} et~al.,}{{Gaudi}
  et~al.}{2018}]{GaudiSeagerMennessonEtAl2018}
{Gaudi} B.~S.,  et~al., 2018, arXiv e-prints, \href
  {https://ui.adsabs.harvard.edu/abs/2018arXiv180909674G} {}

\bibitem[\protect\citeauthoryear{{Give'on}, {Kern}, {Shaklan}, {Moody}  \&
  {Pueyo}}{{Give'on} et~al.}{2007}]{GiveonKernShaklanEtAl2007}
{Give'on} A.,  {Kern} B.,  {Shaklan} S.,  {Moody} D.~C.,   {Pueyo} L.,  2007,
  in Astronomical Adaptive Optics Systems and Applications III. p. 66910A,
  \mn@doi{10.1117/12.733122}

\bibitem[\protect\citeauthoryear{{Groff}}{{Groff}}{2012}]{Groff2012}
{Groff} T.~D.,  2012, PhD thesis, Princeton University

\bibitem[\protect\citeauthoryear{{Guyon}, {Martinache}, {Belikov}  \&
  {Soummer}}{{Guyon} et~al.}{2010a}]{GuyonMartinacheBelikovEtAl2010}
{Guyon} O.,  {Martinache} F.,  {Belikov} R.,   {Soummer} R.,  2010a, \mn@doi
  [\apjs] {10.1088/0067-0049/190/2/220}, \href
  {http://adsabs.harvard.edu/abs/2010ApJS..190..220G} {190, 220}

\bibitem[\protect\citeauthoryear{{Guyon}, {Martinache}, {Garrel}, {Vogt},
  {Yokochi}  \& {Yoshikawa}}{{Guyon}
  et~al.}{2010b}]{GuyonMartinacheGarrelEtAl2010}
{Guyon} O.,  {Martinache} F.,  {Garrel} V.,  {Vogt} F.,  {Yokochi} K.,
  {Yoshikawa} T.,  2010b, in Adaptive Optics Systems II. p. 773624,
  \mn@doi{10.1117/12.857878}

\bibitem[\protect\citeauthoryear{{Guyon}, {Martinache}, {Cady}, {Belikov},
  {Balasubramanian}, {Wilson}, {Clergeon}  \& {Mateen}}{{Guyon}
  et~al.}{2012}]{GuyonMartinacheCadyEtAl2012}
{Guyon} O.,  {Martinache} F.,  {Cady} E.~J.,  {Belikov} R.,  {Balasubramanian}
  K.,  {Wilson} D.,  {Clergeon} C.~S.,   {Mateen} M.,  2012, in Adaptive Optics
  Systems III. p. 84471X, \mn@doi{10.1117/12.927181}

\bibitem[\protect\citeauthoryear{{Guyon}, {Hinz}, {Cady}, {Belikov}  \&
  {Martinache}}{{Guyon} et~al.}{2014}]{GuyonHinzCadyEtAl2014}
{Guyon} O.,  {Hinz} P.~M.,  {Cady} E.,  {Belikov} R.,   {Martinache} F.,  2014,
  \mn@doi [\apj] {10.1088/0004-637X/780/2/171}, \href
  {http://adsabs.harvard.edu/abs/2014ApJ...780..171G} {780, 171}

\bibitem[\protect\citeauthoryear{{Janin-Potiron}, {Martinez}, {Baudoz}  \&
  {Carbillet}}{{Janin-Potiron}
  et~al.}{2016}]{Janin-PotironMartinezBaudozEtAl2016}
{Janin-Potiron} P.,  {Martinez} P.,  {Baudoz} P.,   {Carbillet} M.,  2016,
  \mn@doi [\aap] {10.1051/0004-6361/201628287}, \href
  {https://ui.adsabs.harvard.edu/abs/2016A%26A...592A.110J} {592, A110}

\bibitem[\protect\citeauthoryear{{Janin-Potiron}, {N'Diaye}, {Martinez},
  {Vigan}, {Dohlen}  \& {Carbillet}}{{Janin-Potiron}
  et~al.}{2017}]{Janin-PotironNDiayeMartinezEtAl2017}
{Janin-Potiron} P.,  {N'Diaye} M.,  {Martinez} P.,  {Vigan} A.,  {Dohlen} K.,
  {Carbillet} M.,  2017, \mn@doi [\aap] {10.1051/0004-6361/201730686}, \href
  {https://ui.adsabs.harvard.edu/abs/2017A%26A...603A..23J} {603, A23}

\bibitem[\protect\citeauthoryear{{Jiang}, {Ling}, {Rao}  \& {Shi}}{{Jiang}
  et~al.}{1991}]{JiangLingRaoEtAl1991}
{Jiang} W.,  {Ling} N.,  {Rao} X.,   {Shi} F.,  1991, in {Ealey} M.~A.,  ed.,
  \procspie Vol. 1542, Active and Adaptive Optical Systems. pp 130--137,
  \mn@doi{10.1117/12.48800}

\bibitem[\protect\citeauthoryear{{Jolivet}, {Piron}, {Huby}, {Absil},
  {Delacroix}, {Mawet}, {Surdej}  \& {Habraken}}{{Jolivet}
  et~al.}{2014}]{JolivetPironHubyEtAl2014}
{Jolivet} A.,  {Piron} P.,  {Huby} E.,  {Absil} O.,  {Delacroix} C.,  {Mawet}
  D.,  {Surdej} J.,   {Habraken} S.,  2014, in Advances in Optical and
  Mechanical Technologies for Telescopes and Instrumentation. p. 91515P
  (\mn@eprint {arXiv} {1604.08334}), \mn@doi{10.1117/12.2055791}

\bibitem[\protect\citeauthoryear{{Juanola-Parramon}, {Zimmerman}, {Groff},
  {Pueyo}, {Rizzo}, {Bolcar}  \& {Roberge}}{{Juanola-Parramon}
  et~al.}{2019a}]{JuanolaParramonZimmermanGroffEtAl2019}
{Juanola-Parramon} R.,  {Zimmerman} N.~T.,  {Groff} T.,  {Pueyo} L.,  {Rizzo}
  M.,  {Bolcar} M.,   {Roberge} A.,  2019a, in American Astronomical Society
  Meeting Abstracts \#233. p. 148.06

\bibitem[\protect\citeauthoryear{Juanola-Parramon, Zimmerman, Pueyo, Bolcar,
  Ruane, Krist  \& Groff}{Juanola-Parramon
  et~al.}{2019b}]{JuanolaZimmermanPueyoEtAl2019}
Juanola-Parramon R.,  Zimmerman N.~T.,  Pueyo L.,  Bolcar M.,  Ruane G.,  Krist
  J.,   Groff T.,  2019b, in Techniques and Instrumentation for Detection of
  Exoplanets IX. p. 1111705

\bibitem[\protect\citeauthoryear{Kern et~al.,}{Kern
  et~al.}{2016}]{KernGuyonBelikovEtAl2016}
Kern B.,  et~al., 2016, Journal of Astronomical Telescopes, Instruments, and
  Systems, 2, 011014

\bibitem[\protect\citeauthoryear{Knight, Brewer, Hamilton, Ward, Milster  \&
  Guyon}{Knight et~al.}{2017}]{KnightBrewerHamiltonEtAl2017}
Knight J.~M.,  Brewer J.,  Hamilton R.,  Ward K.,  Milster T.~D.,   Guyon O.,
  2017, in Techniques and Instrumentation for Detection of Exoplanets VIII. p.
  104000N

\bibitem[\protect\citeauthoryear{{Krist}}{{Krist}}{2007}]{Krist2007a}
{Krist} J.~E.,  2007, in Optical Modeling and Performance Predictions III. p.
  66750P, \mn@doi{10.1117/12.731179}

\bibitem[\protect\citeauthoryear{{Krist}, {Belikov}, {Pueyo}, {Mawet}, {Moody},
  {Trauger}  \& {Shaklan}}{{Krist} et~al.}{2011}]{KristBelikovPueyoEtAl2011}
{Krist} J.~E.,  {Belikov} R.,  {Pueyo} L.,  {Mawet} D.~P.,  {Moody} D.,
  {Trauger} J.~T.,   {Shaklan} S.~B.,  2011, in Techniques and Instrumentation
  for Detection of Exoplanets V. pp 81510E--81510E--16,
  \mn@doi{10.1117/12.892772}

\bibitem[\protect\citeauthoryear{{Krist}, {Nemati}  \& {Mennesson}}{{Krist}
  et~al.}{2016}]{KristNematiMennesson2016}
{Krist} J.,  {Nemati} B.,   {Mennesson} B.,  2016, \mn@doi [Journal of
  Astronomical Telescopes, Instruments, and Systems]
  {10.1117/1.JATIS.2.1.011003}, \href
  {https://ui.adsabs.harvard.edu/abs/2016JATIS...2a1003K} {2, 011003}

\bibitem[\protect\citeauthoryear{Krist et~al.,}{Krist
  et~al.}{2019}]{KristMartinKuanEtAl2019}
Krist J.,  et~al., 2019, in Techniques and Instrumentation for Detection of
  Exoplanets IX. p. 1111702

\bibitem[\protect\citeauthoryear{{Kuhn}, {Potter}  \& {Parise}}{{Kuhn}
  et~al.}{2001}]{KuhnPotterParise2001}
{Kuhn} J.~R.,  {Potter} D.,   {Parise} B.,  2001, \mn@doi [\apjl]
  {10.1086/320686}, \href {http://adsabs.harvard.edu/abs/2001ApJ...553L.189K}
  {553, L189}

\bibitem[\protect\citeauthoryear{{Lafreni{\`e}re}, {Marois}, {Doyon}, {Nadeau}
  \& {Artigau}}{{Lafreni{\`e}re} et~al.}{2007}]{LafreniereMaroisDoyonEtAl2007}
{Lafreni{\`e}re} D.,  {Marois} C.,  {Doyon} R.,  {Nadeau} D.,   {Artigau}
  {\'E}.,  2007, \mn@doi [\apj] {10.1086/513180}, \href
  {http://adsabs.harvard.edu/abs/2007ApJ...660..770L} {660, 770}

\bibitem[\protect\citeauthoryear{Lawson et~al.,}{Lawson
  et~al.}{2013}]{LawsonBelikovCashEtAl2013}
Lawson P.~R.,  et~al., 2013, in Techniques and Instrumentation for Detection of
  Exoplanets VI. p. 88641F

\bibitem[\protect\citeauthoryear{{Macintosh} et~al.,}{{Macintosh}
  et~al.}{2007}]{MacintoshGrahamPalmerEtAl2007}
{Macintosh} B.,  et~al., 2007, in American Astronomical Society Meeting
  Abstracts. p.~782

\bibitem[\protect\citeauthoryear{{Marois}, {Doyon}, {Racine}  \&
  {Nadeau}}{{Marois} et~al.}{2000}]{MaroisDoyonRacineEtAl2000}
{Marois} C.,  {Doyon} R.,  {Racine} R.,   {Nadeau} D.,  2000, \mn@doi [\pasp]
  {10.1086/316492}, \href {http://adsabs.harvard.edu/abs/2000PASP..112...91M}
  {112, 91}

\bibitem[\protect\citeauthoryear{{Marois}, {Doyon}, {Racine}, {Nadeau},
  {Lafreniere}, {Vallee}, {Riopel}  \& {Macintosh}}{{Marois}
  et~al.}{2005}]{MaroisDoyonRacineEtAl2005}
{Marois} C.,  {Doyon} R.,  {Racine} R.,  {Nadeau} D.,  {Lafreniere} D.,
  {Vallee} P.,  {Riopel} M.,   {Macintosh} B.,  2005, \jrasc, \href
  {http://adsabs.harvard.edu/abs/2005JRASC..99..130M} {99, 130}

\bibitem[\protect\citeauthoryear{{Marois}, {Lafreni{\`e}re}, {Doyon},
  {Macintosh}  \& {Nadeau}}{{Marois}
  et~al.}{2006a}]{MaroisLafreniereDoyonEtAl2006}
{Marois} C.,  {Lafreni{\`e}re} D.,  {Doyon} R.,  {Macintosh} B.,   {Nadeau} D.,
   2006a, \mn@doi [\apj] {10.1086/500401}, \href
  {http://adsabs.harvard.edu/abs/2006ApJ...641..556M} {641, 556}

\bibitem[\protect\citeauthoryear{{Marois}, {Phillion}  \& {Macintosh}}{{Marois}
  et~al.}{2006b}]{MaroisPhillionMacintosh2006}
{Marois} C.,  {Phillion} D.~W.,   {Macintosh} B.,  2006b, in Society of
  Photo-Optical Instrumentation Engineers (SPIE) Conference Series. p. 62693M
  (\mn@eprint {} {astro-ph/0607002}), \mn@doi{10.1117/12.672263}

\bibitem[\protect\citeauthoryear{Martinache et~al.,}{Martinache
  et~al.}{2014}]{MartinacheGuyonJovanovicEtAl2014}
Martinache F.,  et~al., 2014, Publications of the Astronomical Society of the
  Pacific, 126, 565

\bibitem[\protect\citeauthoryear{{Martinez} et~al.,}{{Martinez}
  et~al.}{2014}]{MartinezPreisGouvretEtAl2014}
{Martinez} P.,  et~al., 2014, in Ground-based and Airborne Telescopes V. p.
  91454E, \mn@doi{10.1117/12.2055338}

\bibitem[\protect\citeauthoryear{Martinez, Beaulieu, Guyon, Abe, Gouvret,
  Dejonghe  \& Preis}{Martinez et~al.}{2018}]{MartinezBeaulieuGuyonEtAl2018}
Martinez P.,  Beaulieu M.,  Guyon O.,  Abe L.,  Gouvret C.,  Dejonghe J.,
  Preis O.,  2018, in Ground-based and Airborne Instrumentation for Astronomy
  VII. p. 1070243

\bibitem[\protect\citeauthoryear{{Martinez} et~al.,}{{Martinez}
  et~al.}{2019}]{MartinezBeaulieuBarjotEtAl2019}
{Martinez} P.,  et~al., {2019}, {\aap, under press}

\bibitem[\protect\citeauthoryear{Matthews, Crepp, Vasisht  \& Cady}{Matthews
  et~al.}{2017}]{MatthewsCreppVasishtEtAl2017}
Matthews C.~T.,  Crepp J.~R.,  Vasisht G.,   Cady E.,  2017, Journal of
  Astronomical Telescopes, Instruments, and Systems, 3, 045001

\bibitem[\protect\citeauthoryear{{Mawet}, {Riaud}, {Absil}  \&
  {Surdej}}{{Mawet} et~al.}{2005}]{MawetRiaudAbsilEtAl2005}
{Mawet} D.,  {Riaud} P.,  {Absil} O.,   {Surdej} J.,  2005, \mn@doi [\apj]
  {10.1086/462409}, \href {http://adsabs.harvard.edu/abs/2005ApJ...633.1191M}
  {633, 1191}

\bibitem[\protect\citeauthoryear{{Mawet} et~al.,}{{Mawet}
  et~al.}{2017}]{MawetRuaneXuanEtAl2017}
{Mawet} D.,  et~al., 2017, \mn@doi [\apj] {10.3847/1538-4357/aa647f}, \href
  {https://ui.adsabs.harvard.edu/abs/2017ApJ...838...92M} {838, 92}

\bibitem[\protect\citeauthoryear{Mazoyer et~al.,}{Mazoyer
  et~al.}{2019}]{MazoyerBaudozBelikovEtAl2019}
Mazoyer J.,  et~al., 2019, Astro2020: Decadal Survey on Astronomy and
  Astrophysics, APC white papers

\bibitem[\protect\citeauthoryear{{N'Diaye} et~al.,}{{N'Diaye}
  et~al.}{2013}]{NDiayeChoquetPueyoEtAl2013}
{N'Diaye} M.,  et~al., 2013, in Techniques and Instrumentation for Detection of
  Exoplanets VI. p. 88641K (\mn@eprint {arXiv} {1407.0979}),
  \mn@doi{10.1117/12.2023718}

\bibitem[\protect\citeauthoryear{{N'Diaye}, Martinache, Jovanovic, Lozi, Guyon,
  Norris, Ceau  \& Mary}{{N'Diaye}
  et~al.}{2018}]{NDiayeMartinacheJovanovicEtAl2018}
{N'Diaye} M.,  Martinache F.,  Jovanovic N.,  Lozi J.,  Guyon O.,  Norris B.,
  Ceau A.,   Mary D.,  2018, Astronomy \& Astrophysics, 610, A18

\bibitem[\protect\citeauthoryear{{Newman}, {Conway}, {Belikov}  \&
  {Guyon}}{{Newman} et~al.}{2016}]{NewmanConwayBelikovEtAl2016}
{Newman} K.,  {Conway} J.,  {Belikov} R.,   {Guyon} O.,  2016, \mn@doi [\pasp]
  {10.1088/1538-3873/128/963/055003}, \href
  {https://ui.adsabs.harvard.edu/abs/2016PASP..128e5003N} {128, 055003}

\bibitem[\protect\citeauthoryear{{Paul}, {Mugnier}, {Sauvage}, {Dohlen}  \&
  {Ferrari}}{{Paul} et~al.}{2013}]{PaulMugnierSauvageEtAl2013}
{Paul} B.,  {Mugnier} L.,  {Sauvage} J.-F.,  {Dohlen} K.,   {Ferrari} M.,
  2013, Optics Express, 21, 31751

\bibitem[\protect\citeauthoryear{{Pueyo} \& {Norman}}{{Pueyo} \&
  {Norman}}{2013}]{PueyoNorman2013a}
{Pueyo} L.,  {Norman} C.,  2013, \mn@doi [\apj] {10.1088/0004-637X/769/2/102},
  \href {http://adsabs.harvard.edu/abs/2013ApJ...769..102P} {769, 102}

\bibitem[\protect\citeauthoryear{Pueyo, Kay, Kasdin, Groff, McElwain,
  Give{\textquotesingle}on  \& Belikov}{Pueyo
  et~al.}{2009}]{PueyoKayKasdinEtAl2009}
Pueyo L.,  Kay J.,  Kasdin N.~J.,  Groff T.,  McElwain M.,
  Give{\textquotesingle}on A.,   Belikov R.,  2009, \mn@doi [Applied Optics]
  {10.1364/ao.48.006296}, 48, 6296

\bibitem[\protect\citeauthoryear{Pueyo et~al.,}{Pueyo
  et~al.}{2019}]{PueyoStarkJuanolaaEtAl2019}
Pueyo L.,  et~al., 2019, in Techniques and Instrumentation for Detection of
  Exoplanets IX. p. 1111703

\bibitem[\protect\citeauthoryear{{Racine}, {Walker}, {Nadeau}, {Doyon}  \&
  {Marois}}{{Racine} et~al.}{1999}]{RacineWalkerNadeauEtAl1999}
{Racine} R.,  {Walker} G.~A.~H.,  {Nadeau} D.,  {Doyon} R.,   {Marois} C.,
  1999, \mn@doi [\pasp] {10.1086/316367}, \href
  {http://adsabs.harvard.edu/abs/1999PASP..111..587R} {111, 587}

\bibitem[\protect\citeauthoryear{{Riggs}, {Ruane}, {Sidick}, {Coker}, {Kern}
  \& {Shaklan}}{{Riggs} et~al.}{2018}]{RiggsRuaneSidickEtAl2018}
{Riggs} A.~J.~E.,  {Ruane} G.,  {Sidick} E.,  {Coker} C.,  {Kern} B.~D.,
  {Shaklan} S.~B.,  2018, in Space Telescopes and Instrumentation 2018:
  Optical, Infrared, and Millimeter Wave. p. 106982V,
  \mn@doi{10.1117/12.2313812}

\bibitem[\protect\citeauthoryear{{Rosenthal}, {Gurwell}  \& {Ho}}{{Rosenthal}
  et~al.}{1996}]{RosenthalGurwellHo1996}
{Rosenthal} E.~D.,  {Gurwell} M.~A.,   {Ho} P.~T.~P.,  1996, \mn@doi [\nat]
  {10.1038/384243a0}, \href {http://adsabs.harvard.edu/abs/1996Natur.384..243R}
  {384, 243}

\bibitem[\protect\citeauthoryear{Sauvage et~al.,}{Sauvage
  et~al.}{2016}]{SauvageFuscoLambEtAl2016}
Sauvage J.-F.,  et~al., 2016, in Adaptive Optics Systems V. p. 990916

\bibitem[\protect\citeauthoryear{Savransky, Macintosh, Thomas, Poyneer, Palmer,
  De~Rosa  \& Hartung}{Savransky
  et~al.}{2012}]{SavranskyMacintoshThomasEtAl2012}
Savransky D.,  Macintosh B.~A.,  Thomas S.~J.,  Poyneer L.~A.,  Palmer D.~W.,
  De~Rosa R.~J.,   Hartung M.,  2012, in Adaptive Optics Systems III. p. 84476S

\bibitem[\protect\citeauthoryear{{Smith}}{{Smith}}{1987}]{Smith1987}
{Smith} W.~H.,  1987, \mn@doi [\pasp] {10.1086/132124}, \href
  {http://adsabs.harvard.edu/abs/1987PASP...99.1344S} {99, 1344}

\bibitem[\protect\citeauthoryear{Soummer et~al.,}{Soummer
  et~al.}{2019}]{SoummerBradyBrooksEtAl2019}
Soummer R.,  et~al., 2019, \mn@doi [Proceedings of SPIE The International
  Society for Optical Engineering] {10.1117/12.2314110}

\bibitem[\protect\citeauthoryear{{Sparks} \& {Ford}}{{Sparks} \&
  {Ford}}{2002}]{SparksFord2002}
{Sparks} W.~B.,  {Ford} H.~C.,  2002, \mn@doi [\apj] {10.1086/342401}, \href
  {http://adsabs.harvard.edu/abs/2002ApJ...578..543S} {578, 543}

\bibitem[\protect\citeauthoryear{{Trauger}, {Moody}, {Gordon}, {Krist}  \&
  {Mawet}}{{Trauger} et~al.}{2011}]{TraugerMoodyGordonEtAl2011}
{Trauger} J.,  {Moody} D.,  {Gordon} B.,  {Krist} J.,   {Mawet} D.,  2011, in
  Techniques and Instrumentation for Detection of Exoplanets V. p. 81510G,
  \mn@doi{10.1117/12.895032}

\makeatother
\end{thebibliography}



\bsp	
\label{lastpage}
\end{document}